\documentclass[preprint,preprintnumbers,aps,superscriptaddress,nofootinbib,tightenlines,floatfix]{revtex4}

\usepackage{graphicx}
\usepackage{bm}
\usepackage{comment}
\usepackage{color}

\newcommand{\gsim}{\raisebox{-0.7ex}{$\stackrel{\textstyle >}{\sim}$ }}
\newcommand{\lsim}{\raisebox{-0.7ex}{$\stackrel{\textstyle <}{\sim}$ }}
\def\si{^1 \hskip -0.03in S _0}
\def\siii{^3 \hskip -0.025in S _1}
\def\diii{^3 \hskip -0.03in D _1}

\newcount\hour \newcount\hourminute \newcount\minute 
\hour=\time \divide \hour by 60
\hourminute=\hour \multiply \hourminute by 60
\minute=\time \advance \minute by -\hourminute
\newcommand{\mydate}{\ \today \ - \number\hour :\number\minute}

\begin{document}

\begin{figure}[!t]
  \vskip -1.5cm
  \leftline{\includegraphics[width=0.25\textwidth]{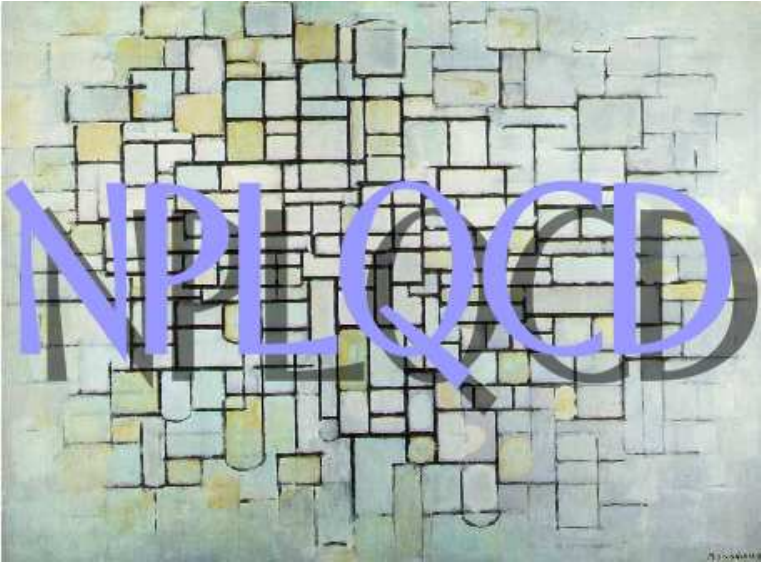}}
\end{figure}

\preprint{
\vbox{ 
\hbox{ICCUB-11-165} 
\hbox{NT@UW-11-21}
\hbox{NT-LBNL-11-017}
\hbox{UCB-NPAT-11-012} 
\hbox{UNH-11-5} 
}}

\title{
The Deuteron and Exotic Two-Body Bound States \\
from Lattice QCD
}

\author{S.R.~Beane} 
\affiliation{Department of Physics, University
  of New Hampshire, Durham, NH 03824-3568, USA}

\author{E.~Chang}
\affiliation{Dept. d'Estructura i Constituents de la Mat\`eria. 
Institut de Ci\`encies del Cosmos (ICC),
Universitat de Barcelona, Mart\'{\i} i Franqu\`es 1, E08028-Spain}

\author{W.~Detmold} 
\affiliation{Department of Physics, College of William and Mary, Williamsburg,
  VA 23187-8795, USA}
\affiliation{Jefferson Laboratory, 12000 Jefferson Avenue, 
Newport News, VA 23606, USA}

\author{H.W.~Lin}
\affiliation{Department of Physics,
  University of Washington, Box 351560, Seattle, WA 98195, USA}

\author{T.C.~Luu}
\affiliation{N Division, Lawrence Livermore National Laboratory, Livermore, CA
  94551, USA}

\author{K.~Orginos}
\affiliation{Department of Physics, College of William and Mary, Williamsburg,
  VA 23187-8795, USA}
\affiliation{Jefferson Laboratory, 12000 Jefferson Avenue, 
Newport News, VA 23606, USA}

\author{A.~Parre\~no}
\affiliation{Dept. d'Estructura i Constituents de la Mat\`eria. 
Institut de Ci\`encies del Cosmos (ICC),
Universitat de Barcelona, Mart\'{\i} i Franqu\`es 1, E08028-Spain}

\author{M.J.~Savage} \affiliation{Department of Physics,
  University of Washington, Box 351560, Seattle, WA 98195, USA}

\author{A.~Torok} \affiliation{Department of Physics, Indiana University,
  Bloomington, IN 47405, USA}

\author{A.~Walker-Loud}
\affiliation{Lawrence Berkeley National Laboratory, Berkeley, CA 94720, USA}

\collaboration{ NPLQCD Collaboration }

\date{\mydate}

\begin{abstract}
  \noindent 
Results of a high-statistics, multi-volume Lattice QCD exploration of
the deuteron, the di-neutron, the H-dibaryon, and the $\Xi^-\Xi^-$ system
at a pion mass of $m_\pi\sim 390~{\rm MeV}$ are presented.
 Calculations were performed with an anisotropic
$n_f=2+1$ Clover discretization in four
  lattice volumes of spatial extent $L\sim 2.0, 2.5, 3.0$ and
  $4.0~{\rm fm}$, with a lattice spacing of $b_s\sim
  0.123~{\rm fm}$ in the spatial-direction, and $b_t \sim b_s/3.5$ in the
  time-direction.
The $\Xi^-\Xi^- $ is found to be bound  by  $B_{\Xi^-\Xi^-}=14.0(1.4)(6.7)~{\rm MeV}$, 
consistent with expectations based upon phenomenological models and low-energy
effective field theories
constrained by nucleon-nucleon and hyperon-nucleon scattering data
at the physical light-quark masses. We find weak evidence that
both the deuteron and the di-neutron are bound at this pion mass,
with binding energies of 
$B_d=11(05)(12)~{\rm MeV}$ and 
$B_{nn}=7.1(5.2)(7.3)~{\rm MeV}$, respectively. 
With an increased number of measurements and a refined analysis, 
the binding energy of the H-dibaryon is $B_H=13.2(1.8)(4.0)~{\rm MeV}$ at this
pion mass,
updating our previous result.
\end{abstract}
\maketitle
\tableofcontents
\vfill\eject
%

%%%%%%%%%%%%%%%%%%%%%%%%%%%%%%%%%%%%%%%%%%%%%%%%%%%%
\section{Introduction}
\label{sec:intro}
\noindent
A major objective for nuclear physicists is to 
establish the technology with which
to reliably calculate the
properties and interactions of nuclei
and to be able to quantify the
uncertainties in such calculations.
Achieving this objective
will have broad impact, from establishing the behavior of matter under
extreme conditions such as those that arise in the interior of neutron stars,
to refining predictions for  the array of  isotopes produced in nuclear
reactors, and even to answering anthropic questions about the nature of our universe.
While nuclear phenomenology generally describes experimentally measured
quantities, its ability to make high precision
and accurate predictions for quantities that 
cannot be accessed experimentally
is limited.
This situation is on the verge of dramatically improving.
The underlying theory of the strong interactions is known to be quantum
chromodynamics (QCD), and 
the computational resources 
now available are beginning to allow for {\it ab initio} calculations of basic
quantities in nuclear physics.
With further increases in computational power and advances in algorithms, this
trend will continue and our understanding of, and our ability to calculate,
light and exotic nuclei will be placed on  a solid foundation.

In nature, two nucleons in the $\siii-\diii$ coupled channels bind to form the 
simplest nucleus, the deuteron ($J^\pi=1^+$), 
with a binding energy of $B_d = 2.224644(34)~{\rm MeV}$,
and nearly bind into a di-neutron in the $\si$ channel.
However, little is known experimentally about possible bound states in more
exotic channels, for instance those containing strange quarks.  
The most famous exotic channel that has been postulated to support a bound
state (the  H-dibaryon~\cite{Jaffe:1976yi})
has the quantum numbers of $\Lambda\Lambda$ (total angular momentum
$J^\pi=0^+$, isospin $I=0$ and strangeness $s=-2$).  
In this channel, all six quarks in naive quark models, like the MIT bag model,
can be in the lowest-energy single-particle state.
Additionally, more extensive analyses using one-boson-exchange (OBE) models~\cite{Stoks:1999bz} and
low-energy effective field theories
(EFT)~\cite{Miller:2006jf,Haidenbauer:2009qn}, 
both constrained by experimentally
measured  nucleon-nucleon (NN) and  hyperon-nucleon (YN) cross-sections 
and the approximate SU(3) flavor symmetry of
the strong interactions, suggest that other exotic channels also support
bound states.  
In the limit of SU(3) flavor symmetry, the $\si$-channels
are in symmetric irreducible representations of
${\bf 8}\otimes {\bf 8} = {\bf 27}\oplus {\bf 10}\oplus \overline{{\bf 10}}
\oplus {\bf 8} \oplus {\bf 8} \oplus {\bf 1}$, and hence the $\Xi^-\Xi^-$,
$\Sigma^-\Sigma^-$, and $nn$ 
(along with $n\Sigma^-$ and $\Sigma^-\Xi^-$)
all transform in the ${\bf 27}$.
YN and NN scattering data along with the leading SU(3) breaking effects, arising
from the light-meson and baryon masses, suggest that $\Xi^-\Xi^-$  and
$\Sigma^-\Sigma^-$  are bound at the physical values of the light-quark 
masses~\cite{Stoks:1999bz,Miller:2006jf,Haidenbauer:2009qn}.

Recently, the first steps have been taken
towards calculating the binding energies of light
nuclei directly from QCD.
Early exploratory quenched calculations of the NN scattering 
lengths~\cite{Fukugita:1994na,Fukugita:1994ve} 
performed more than a decade ago have been 
superseded by
$n_f=2+1$ calculations within the last few
years~\cite{Beane:2006mx,Beane:2009py}
(and added to by further quenched
calculations~\cite{Aoki:2008hh,Aoki:2009ji}~\footnote{The HALQCD collaboration
  has produced non-local, energy-dependent, and sink-operator dependent quantities 
from lattice spatial correlation functions 
that contain the same, but no more,
information than the NN energy eigenvalues in the lattice volume(s), e.g. Ref.~\cite{Ishii:2006ec}.
}).
Further, the first quenched calculations of the deuteron~\cite{Yamazaki:2011nd}, 
$^3$He and $^4$He~\cite{Yamazaki:2009ua} have
been performed, 
along with $n_f=2+1$ calculations of $^3$He~\cite{Beane:2009gs} and multi-baryon systems
containing strange quarks~\cite{Beane:2009gs}.
Efforts to explore nuclei and nuclear matter using the strong coupling limit of
QCD 
have led to some interesting observations~\cite{deForcrand:2009dh}.
Recently, $n_f=2+1$ calculations by us (NPLQCD)~\cite{Beane:2010hg},  and subsequent $n_f=3$
calculations by the HALQCD collaboration~\cite{Inoue:2010es}, 
have provided evidence that the H-dibaryon (with the quantum numbers of
$\Lambda\Lambda$) is bound at a pion mass of $ m_\pi\sim 390~{\rm MeV}$ [NPLQCD]
and
$m_\pi\sim 837~{\rm MeV}$ [HALQCD]~\footnote{One should note that both
  calculations were performed at approximately the same spatial lattice spacing
of $b\sim 0.12~{\rm fm}$.}.  
Extrapolations to the physical
light-quark masses suggest a weakly bound H-dibaryon or a near threshold
resonance exists in this channel~\cite{Beane:2011xf,Shanahan:2011su}.

In this work,
which is a continuation of our high-statistics Lattice QCD
explorations~\cite{Beane:2009kya,Beane:2009gs,Beane:2009py,Beane:2011pc},
we present evidence for 
$\Xi^-\Xi^- (\si)$ and  
H-dibaryon (refining  of our results presented in Ref.~\cite{Beane:2010hg})
bound states, and   weak evidence 
for a bound deuteron and di-neutron at a pion mass of $m_\pi\sim 390~{\rm MeV}$.
The results were obtained from four ensembles of $n_f=2+1$
anisotropic clover gauge-field configurations 
with a spatial lattice spacing of $b_s\sim
0.123~{\rm fm}$, an anisotropy of $\xi\sim3.5$ and with cubic volumes
of spatial extent $L\sim 2.0, 2.5, 3.0$ and $4.0~{\rm fm}$.

In section~\ref{sec:Method}, a concise description of the specific LQCD
technology and 
computational details
relevant to the present two-body bound state calculations
are given.
Section~\ref{sec:B} presents the results of the LQCD calculations of the single
baryon masses and dispersion relations (critical for understanding bound
systems), 
and in section~\ref{sec:BB} the results
for the bound states are presented.  
Discussions and our
conclusions can be found
in section~\ref{sec:Conclusions}.

%%%%%%%%%%%%%%%%%%%%%%%%%%%%%%%%%%%%%%%%%%%%%%%%%%%%
\section{Lattice QCD Calculations}
\label{sec:Method}
\noindent
Lattice QCD (LQCD) is a technique in which space-time is discretized into a
four-dimensional grid
and the QCD path integral over the quark and gluon fields at each
point in the grid is performed in Euclidean space-time
using Monte Carlo methods.
A LQCD calculation of a given quantity will differ from its actual value 
because of the finite volume of the space-time (with $L^3\times T$ lattice points)
over which the fields exist, and
the finite separation between space-time points (the lattice-spacing).
However, such deviations can be systematically removed by performing
calculations in multiple volumes with multiple lattice spacings, and
extrapolating using the
theoretically known functional dependences on each.
In the following subsections, we review the details of LQCD calculations
relevant  to the current work and introduce the ensembles studied herein.

%%%%%%%%%%%%%%%%%%%%%%%%%%%%%%%%%%%%%%%%%%%%%%%%%%%%
\subsection{L\"uscher's Method for Two-Body Systems Including Bound States}
\label{sec:Luscher}
\noindent
The hadron-hadron scattering amplitude below the inelastic threshold can be
determined from two-hadron energy levels in the lattice volume using 
L\"uscher's method~\cite{Hamber:1983vu,Luscher:1986pf,Luscher:1990ux}.
In the situation where only a single
scattering channel is kinematically allowed, the deviation of the
energy eigenvalues of the two-hadron system in the lattice volume from
the sum of the single-hadron energies is related to the scattering phase
shift, $\delta (k)$,  at the measured two-hadron energies.  
For energy eigenvalues above kinematic thresholds
where multiple channels contribute, a coupled-channels analysis is
required as a single phase shift does not parameterize the S-matrix.
Such analyses can be performed, but they are  not required in the current context.
The energy shift for two particles $A$ and $B$, $\Delta E = E_{AB} -
E_A - E_B$, can be determined from the correlation functions for
systems containing one and two hadrons.  For baryon-baryon systems,
correlation functions of the form
\begin{eqnarray}
  \label{eq:correlators}
  C_{{\cal B};\Gamma}({\bf p},t) &=& \sum_{\bf x} \ e^{i{\bf p}\cdot{\bf x}} \ 
  \Gamma_{\alpha}^\beta\ 
  \langle {\cal B}_\alpha({\bf x},t)\  \overline{\cal
    B}_\beta({\bf x}_0,0)\rangle
  \\
  C_{{\cal B}_1,{\cal B}_2;\Gamma}({\bf p}_1,{\bf p}_2,t) 
  &=& \sum_{{\bf x}_1,{\bf
      x}_2} e^{i{\bf p}_1\cdot{\bf x}_1} 
  e^{i{\bf p}_2\cdot{\bf x}_2} 
  \Gamma_{\beta_1\beta_2}^{\alpha_1\alpha_2}
  \langle {\cal B}_{1,\alpha_1}({\bf x}_1,t){\cal B}_{2,\alpha_2}({\bf x}_2,t) 
  \overline{\cal
    B}_{1,\beta_1}({\bf x}_0,0) \overline{\cal
    B}_{2,\beta_2}({\bf x}_0,0) \rangle \,,\nonumber 
\end{eqnarray}
are used, where ${\cal B}$ denotes a baryon interpolating operator,
$\alpha_i$ and $\beta_i$ are Dirac indices, 
and the $\Gamma$ are Dirac matrices that typically project onto
particular parity and  angular momentum states.  
The $\langle ... \rangle$ denote averaging over the gauge-field configurations
and ${\bf x}_0$ is the location of the source.
The interpolating operators 
are only constrained by the quantum numbers of the system of interest, and
the simplest forms are 
\begin{eqnarray}
  \label{eq:7}
  p_\alpha({\bf x},t) &=& \epsilon^{ijk} u_\alpha^i({\bf
    x},t)\left[ u^{j {\sf T}}({\bf x},t)C\gamma_5 d^k({\bf
      x},t)\right]\,,
  \nonumber \\
  \Lambda_\alpha({\bf x},t) &=& \epsilon^{ijk} s_\alpha^i({\bf
    x},t)\left[ u^{j {\sf T}}({\bf x},t)C\gamma_5 d^k({\bf x},t)\right]\,,
  \nonumber \\
  \Sigma^+_\alpha({\bf x},t) &=& \epsilon^{ijk} u_\alpha^i({\bf
    x},t)\left[u^{j {\sf T}}({\bf x},t)C\gamma_5 s^k({\bf x},t)\right]\,,
  \nonumber \\
  \Xi^0_\alpha({\bf x},t) &=& \epsilon^{ijk} s_\alpha^i({\bf
    x},t)\left[ u^{j {\sf T}}({\bf x},t)C\gamma_5 s^k({\bf x},t)\right]\,,
\end{eqnarray}
where $C$ is the charge-conjugation matrix and $ijk$ are color
indices. Other hadrons in the lowest-lying octet can be obtained from
the appropriate combinations of quark flavors.  The brackets in the
interpolating operators indicate contraction of spin indices into a
spin-0 ``diquark''.  
Away from the time slice of the source (in this case
$t=0$), these correlation functions behave as 
\begin{eqnarray}
  \label{eq:5}
  C_{{\cal H}_A}^{(i,f)}({\bf p},t)
  & = & 
  \sum_n \ Z_{n;A}^{(i)}({\bf p}) \ Z_{n;A}^{(f)}({\bf p}) 
  \ e^{- E_n^{(A)}({\bf p})\  t} \,, \\
  C_{{\cal H}_A{\cal H}_{B}}^{(i,f)}({\bf p},-{\bf p},t)
  & = &  \sum_n\  Z_{n;AB}^{(i)}({\bf p}) \  Z_{n;AB}^{(f)}({\bf p}) 
  \ e^{- E_{n}^{(AB)}({\bf 0})\  t} \,,
\end{eqnarray}
where $E_0^{(A)}({\bf 0})=m_A$ and $E_n^{(AB)}({\bf 0})$ are the
energy eigenvalues of the two-hadron system at rest  in
the lattice volume. 
The quantities $Z^{(i)}_{n;X}$ ($Z^{(f)}_{n;X}$) are determined by the overlap of the source
(sink) onto the $n^{\rm th}$ energy eigenstate with the quantum numbers of $X$.
At large times, the ratio
\begin{eqnarray}
  \label{eq:6}
  \frac{C_{{\cal H}_A{\cal H}_{B}}^{(i,f)}({\bf p},-{\bf p},t)}{C_{{\cal H}_A}^{(i,f)}({\bf
      0},t)C_{{\cal H}_{B}}^{(i,f)}({\bf 0},t)}
  &\stackrel{t\to\infty}{\longrightarrow}&  
  \widetilde{Z}_{0,AB}^{(i)}({\bf p})\widetilde{Z}_{0,AB}^{(f)}({\bf p})
  \ e^{- \Delta E_{0}^{(AB)}({\bf 0})\    t}
  \,
\end{eqnarray}
decays as a single exponential in time with the energy shift, $\Delta
E_{0}^{(AB)}({\bf 0})$.  In what follows, only the case ${\bf p}={\bf
  0}$ is considered.  
The energy shift of the $n^{\rm th}$ two-hadron state,
\begin{eqnarray}
  \label{eq:3}
  \Delta E_n^{(AB)} 
  &\equiv& E_n^{(AB)}({\bf 0}) - m_A - m_B = \sqrt{k_n^2 + m_A^2} +
  \sqrt{k_n^2 + m_B^2} -m_A -m_B
\ \ ,
\end{eqnarray}
determines a squared momentum, $k_n^2$ (which can
be either positive or negative). 
Below inelastic thresholds, this is related to the real part of
the inverse scattering amplitude via~\footnote{Calculations performed
  on anisotropic lattices require a modified energy-momentum relation,
  and, as a result, eq.~(\ref{eq:3}) becomes
  \begin{eqnarray}
    \label{eq:3xi}
    \Delta E_n^{(AB)} &\equiv& E_n^{(AB)} - m_A - m_B = 
    \sqrt{k_n^2/\xi_A^2 + m_A^2} +
    \sqrt{k_n^2/\xi_B^2 + m_B^2} -m_A -m_B
    \ \ ,
\nonumber
  \end{eqnarray}
  where $\xi_{A,B}$ are the anisotropy factors for particle $A$ and
  particle $B$, respectively, determined from the appropriate
  energy-momentum dispersion relation.  The masses and energy
  splitting are given in terms of temporal lattice units and $k_n$ is
  given in spatial lattice units.    }
\begin{equation}
  k_n\, \cot \delta(k_n) = 
  \frac{1}{\pi\ L} S\left(k_n^2 \left(\frac{L}{2\pi}\right)^2\right)\,,
  \label{eq:Luscher}
\end{equation}
where
\begin{equation}
  \label{eq:Sfun}
  S(x)=\lim_{\Lambda\to\infty} \sum_{\bf j}^{|{\bf
      j}|<\Lambda}\frac{1}{|{\bf j}|^2 - x}  -4\pi\ \Lambda\,,
\end{equation}
thereby implicitly determining the value of the phase shift at the energy 
$\Delta E_n^{(AB)}$ (or the momentum of each particle in the center of momentum
(CoM) frame, $k_n$), 
$\delta(k_n)$~\cite{Hamber:1983vu,Luscher:1986pf,Luscher:1990ux,Luscher:1985dn,Beane:2003da}.  
Thus, the
function $k\cot\delta$ that determines the low-energy elastic-scattering
cross-section, ${\cal A}(k)\propto(k\cot\delta(k)-i\,k)^{-1}$, is
determined at the energy $\Delta E_n^{(AB)}$.

In a channel for which 
one pion exchange (OPE)
is allowed by spin and isospin
considerations, the function $k\cot\delta(k)$ is an analytic function
of $|{\bf k}|^2$ for $|{\bf k}|\leq m_\pi/2$, as 
determined by the $t$-channel cut in the scattering amplitude.  
In this kinematic regime,  
$k\cot\delta(k)$ can be 
expressed in terms of an
effective range expansion (ERE) of the form
\begin{eqnarray}
  k\cot\delta(k) & = & 
  -{1\over a}\ +\ {1\over 2}\  r_0\ |{\bf k}|^2\ +\ ...
  \ \ \ ,
  \label{eq:ere}
\end{eqnarray} 
where $a$ is the scattering length (with the nuclear physics sign
convention) and $r_0$ is the effective range.  
While the magnitude of
the effective range (and higher terms) is set by the pion mass, the
scattering length is unconstrained.   For scattering processes where
OPE does not contribute, the radius of convergence of the ERE  of $k\cot\delta$
is set by the lightest intermediate state in the $t$-channel 
(or by the inelastic threshold).

In the situation where a channel supports a bound state, 
the energy of the bound state at rest
is determined by eq.~(\ref{eq:Luscher}).
For  $k_{-1}^2<0$, and 
setting $k_{-1} = i\kappa$, eq.~(\ref{eq:Luscher}) becomes 
\begin{eqnarray}
k\cot\delta(k)\big|_{k=i\kappa}
\ +\ \kappa
 & = & 
{1\over L}\ 
\sum_{ {\bf m}\ne {\bf 0} }\
{1\over |{\bf m}|}\ 
e^{- |{\bf m}| \kappa L  }
\ =\ 
{1\over L} \ F^{({\bf 0})}(\kappa L)
\ \ \ ,
\label{eq:pcotkappa}
\end{eqnarray}
where
\begin{eqnarray}
F^{({\bf 0})}(\kappa L) 
& = & 
6 \ e^{-\kappa L }
\ +\ 
6 \sqrt{2}\ e^{-\sqrt{2}\kappa L }
\ +\ 
{8\over\sqrt{3}}\ e^{-\sqrt{3}\kappa L }
\ +\ ...
\ \ \ .
\label{eq:Fdef}
\end{eqnarray}
Perturbation theory can be used 
to solve eq.~(\ref{eq:pcotkappa}) when the extent of the volume is much larger
than the size of the bound system, giving~\cite{Luscher:1985dn,Beane:2003da}
\begin{eqnarray}
\kappa & = & 
\kappa_0 \ +\ 
{Z_\psi^2\over L} F^{({\bf 0})}(\kappa_0 L) \ +\ 
{\cal O}\left(e^{-2 \kappa_0 L }/L\right)
\ \ \ {\rm with} \ \ \ 
Z_\psi \ =\ {1\over\sqrt{1 - 2 \kappa_0 {d\over dk^2} k\cot\delta
    \big|_{i\kappa_0}}}
\ .
\label{eq:kappaPert}
\end{eqnarray}
$\kappa_0$ is the solution to 
\begin{eqnarray}
k\cot\delta(k)\big|_{k=i\kappa_0}
\ +\ \kappa_0
 & = & 
0
\ \ \ ,
\end{eqnarray}
which recovers $\cot\delta(k)\big|_{k=i\kappa_0} = +i$, 
and is the infinite-volume binding momentum of the system. 
This analysis has recently been extended to bound systems that are moving in
the lattice volume~\cite{Bour:2011ef,Davoudi:2011md}.

%%%%%%%%%%%%%%%%%%%%%%%%%%%%%%%%%%%%%%%%%%%%%%%%%%%%
\subsection{Computational Overview}
\label{sec:calcs}
\noindent
Anisotropic gauge field configurations have proven useful
for the study of hadronic spectroscopy, and as the calculations
required for studying multi-hadron systems 
rely heavily on spectroscopy,
considerable effort has been put into calculations with clover-improved
Wilson fermion actions with an anisotropic discretization.  
In particular, the $n_f=2+1$ flavor anisotropic Clover Wilson
action~\cite{Okamoto:2001jb,Chen:2000ej} with two steps of stout-link
smearing~\cite{Morningstar:2003gk} of the spatial gauge fields in the
fermion action with a smearing weight of $\rho=0.14$ has been 
used~\cite{Lin:2008pr,Edwards:2008ja}.
The gauge fields entering the fermion action are not smeared in the
time direction, thus preserving the ultra-locality of the action in
the time direction.  Further, a tree-level tadpole-improved Symanzik
gauge action without a  $1\times 2$ rectangle in the time direction is
used. Anisotropy allows for a better extraction of the excited states
as well as additional confidence that plateaus in the effective mass
plots (EMPs) formed from the correlation functions have been observed,
significantly reducing the systematic uncertainties.  
The gauge field generation was performed  by the Hadron Spectrum
Collaboration (HSC) and by us, and these gauge field configurations have been extensively
used for excited hadron spectrum calculations by
HSC~\cite{Dudek:2009qf,Bulava:2010yg,Dudek:2011tt,Morningstar:2011ka,Edwards:2011jj,Lin:2011da}.

The present calculations are performed on four ensembles of gauge
configurations with $L^3\times T$ of $16^3\times 128$, $20^3\times
128$, $24^3\times 128$ and $32^3\times 256$ lattice sites, with an
anisotropy of $b_t=b_s/\xi$ with $\xi\sim 3.5$.  The spatial lattice
spacing of each ensemble is $b_s\sim 0.1227\pm 0.008~{\rm fm}$, giving
spatial lattice extents of $L\sim 2.0, 2.5, 3.0$ and $4.0~{\rm fm}$
respectively.  The same input light-quark mass parameters, $b_t m_l =
-0.0840$ and $b_t m_s = -0.0743$, are used in the production of each
ensemble, giving a pion mass of $m_\pi\sim 390~{\rm MeV}$.  The 
relevant quantities to assign to each ensemble that determine the impact
of the finite lattice volume are $m_\pi L$ and $m_\pi T$, which for
the four ensembles are $m_\pi L \sim 3.86, 4.82, 5.79$ and $7.71$
respectively, and $m_\pi T \sim 8.82, 8.82, 8.82$ and $17.64$.

For the four lattice ensembles, multiple light-quark propagators were
calculated on each configuration. The source location was chosen
randomly in order to minimize correlations among propagators.
On the $\{ 16^3\times 128$, $20^3\times 128$, $24^3\times 128$,
$32^3\times 256\}$ ensembles, an average of $\{224$, $364$, $178$, $174\}$
propagators were calculated on each of $\{2001$, $1195$, $2215$,
$739\}$ gauge field configurations, to give a total number of $\sim
\{4.5$, $4.3$, $3.9$, $1.3 \}\times 10^5$ 
light-quark propagators, respectively.

%%%%%%%%%%%%%%%%%%%%%%%%%%%%%%%%%%%%%%%%%%%%%%%%%%%%
\section{Baryons and Their Dispersion Relations}
\label{sec:B}
\noindent
The single hadron masses calculated  in the four different lattice
volumes are given in Table~\ref{tab:LQCDbaryonmasses}.
Detailed discussions of
the fitting methods used 
in the analysis of the correlation functions
are given in Ref.~\cite{Beane:2009kya,Beane:2009gs,Beane:2009py,Beane:2010em}.
\begin{table}[!ht]
  \caption{Results from the Lattice QCD calculations in  
four lattice volumes with a pion mass of $m_\pi\sim 390~{\rm MeV}$,
a spatial lattice spacing of $b_s\sim 0.123~{\rm fm}$, and with an 
anisotropy factor of $\xi\sim 3.5$.
Infinite-volume extrapolations~\protect\cite{Beane:2011pc} are shown in the
right column.
The masses are in temporal lattice units (t.l.u).
  }
  \label{tab:LQCDbaryonmasses}
  \begin{ruledtabular}
    \begin{tabular}{c||cccc||c}
      $L^3\times T$  &  $16^3\times 128$ &  $20^3\times 128$ &  $24^3\times 128$ &
      $32^3\times 256$  & Extrapolation \\
      \hline
      $L~({\rm fm})$ & $\sim$ 2.0 &  $\sim$2.5 &  $\sim$3.0 &  $\sim$4.0 &
      $\infty$ \\
      $m_\pi L$ & 3.86 & 4.82 & 5.79 & 7.71 &  $\infty$ \\
      $m_\pi T$ & 8.82 & 8.82 & 8.82 & 17.64 &  $\infty$\\
\hline
      $M_N$ (t.l.u.) &  0.21004(44)(85) & 0.20682(34)(45) & 0.20463(27)(36) &
      0.20457(25)(38) & 0.20455(19)(17)\\
      $M_\Lambda$ (t.l.u.)  &  0.22446(45)(78) & 0.22246(27)(38) &
      0.22074(20)(42) & 0.22054(23)(31)& 0.22064(15)(19)\\
      $M_\Sigma$ (t.l.u.)  & 0.22861(38)(67) & 0.22752(32)(43) &
      0.22791(24)(31) & 0.22726(24)(43)  & 0.22747(17)(19)\\
      $M_\Xi$ (t.l.u.) & 0.24192(38)(63) & 0.24101(27)(38) & 0.23975(20)(32) &
      0.23974(17)(31) & 0.23978(12)(18)\\
    \end{tabular}
  \end{ruledtabular}
\end{table}
Infinite volume extrapolations of the results obtained from the four ensembles were
performed in Ref.~\cite{Beane:2011pc}, and are shown in the right-most column in
Table~\ref{tab:LQCDbaryonmasses}.
In physical units, the extrapolated baryon masses are
$M_N = 1151.3(1.1)(1.0)(7.5)~{\rm MeV}$,
$M_\Lambda = 1241.9(0.8)(1.1)(8.1)~{\rm MeV}$,
$M_\Sigma = 1280.3(1.0)(1.1)(8.3)~{\rm MeV}$,
and 
$M_\Xi = 1349.6(0.7)(1.0)(8.8)~{\rm MeV}$~\cite{Beane:2011pc}.
The difference between a mass calculated  in a finite lattice volume and its
infinite-volume extrapolation is due to contributions of the form $\sim e^{-m_\pi
  L}$.  
Such deviations must be small compared to the two-body binding energies 
to ensure that the 
finite volume bindings are due to the T-matrix~\cite{Bedaque:2006yi,Sato:2007ms} 
and not from finite volume distortions of the forces.
It has been shown~\cite{Beane:2010hg,Beane:2011pc} that the largest two volumes, the $24^3\times 128$ and
$32^3\times 256$ ensembles, are sufficiently large to render the 
$\sim e^{-m_\pi L}$ modifications to L\"uscher's eigenvalue relation negligible
at the level of precision we are currently able to achieve.  
In what follows,  we only consider results from these ensembles.

\begin{figure}[!ht]
  \centering
     \includegraphics[width=0.49\textwidth]{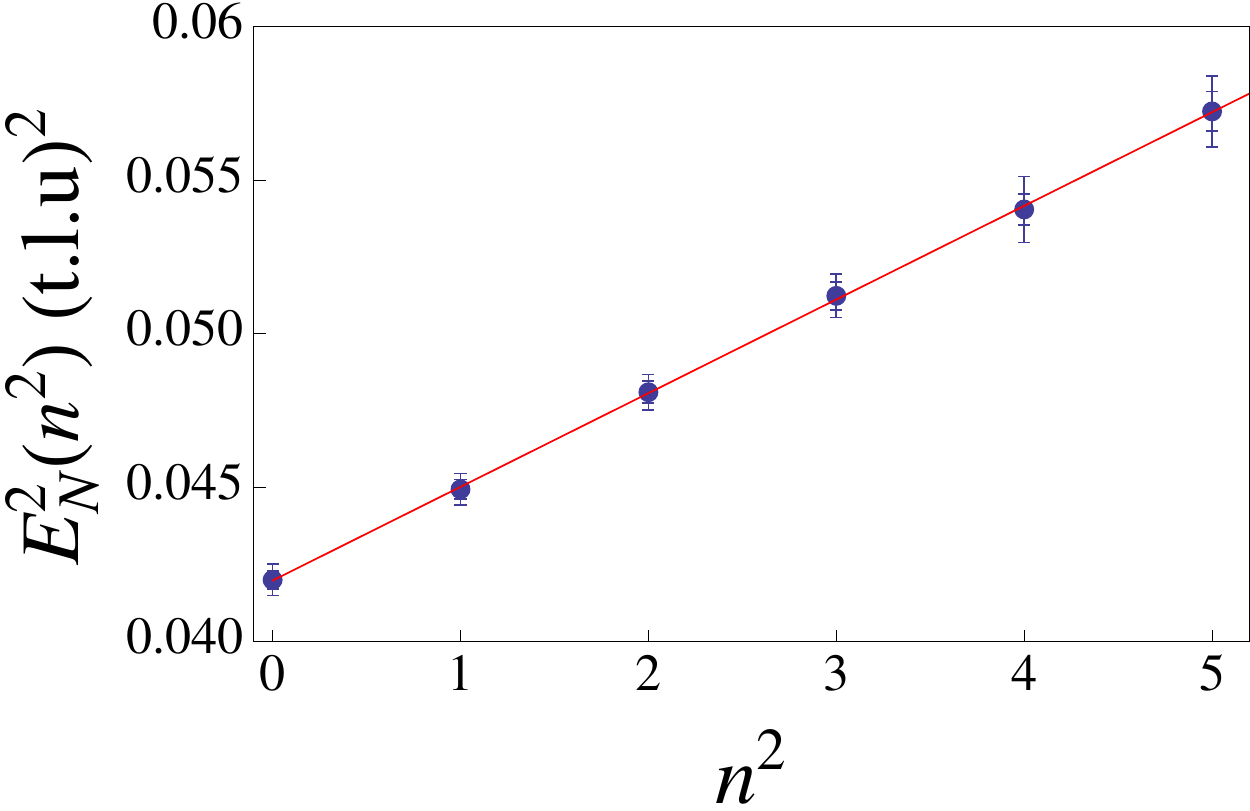}\ \  
     \includegraphics[width=0.49\textwidth]{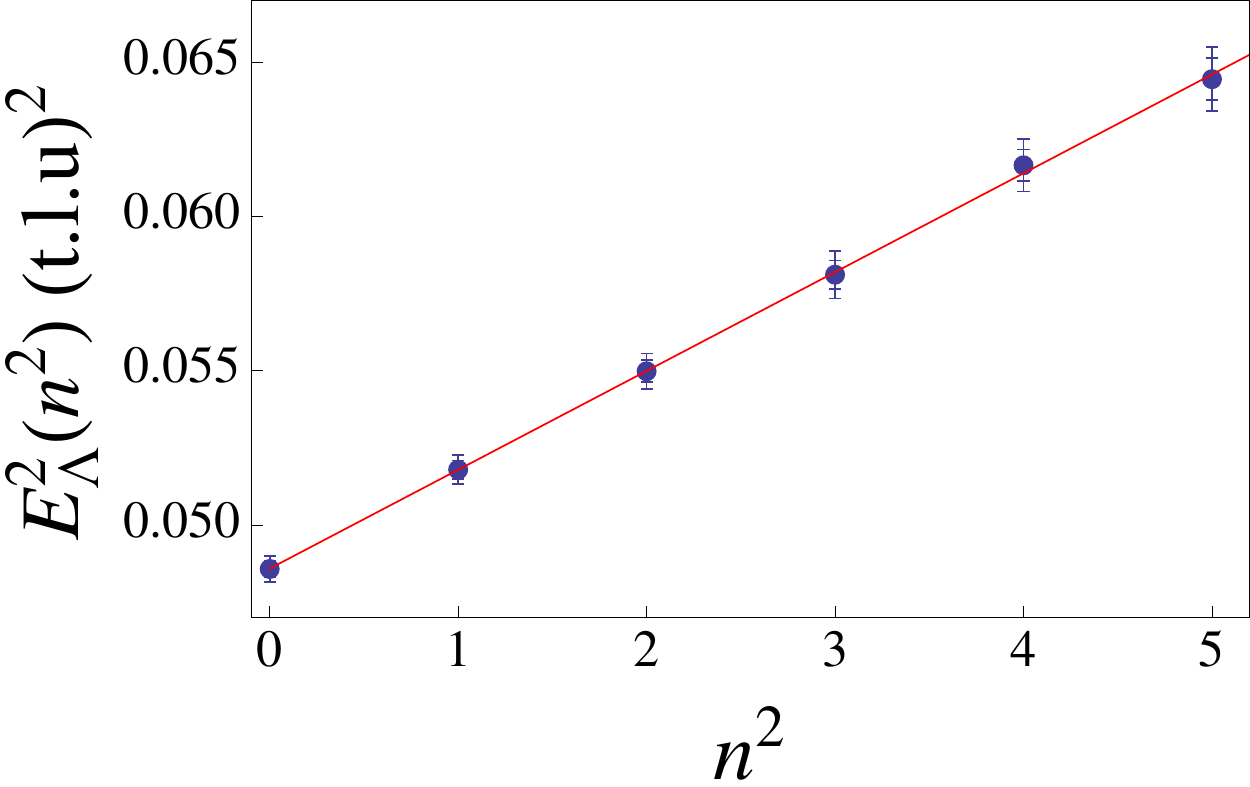}\ \ 
     \includegraphics[width=0.49\textwidth]{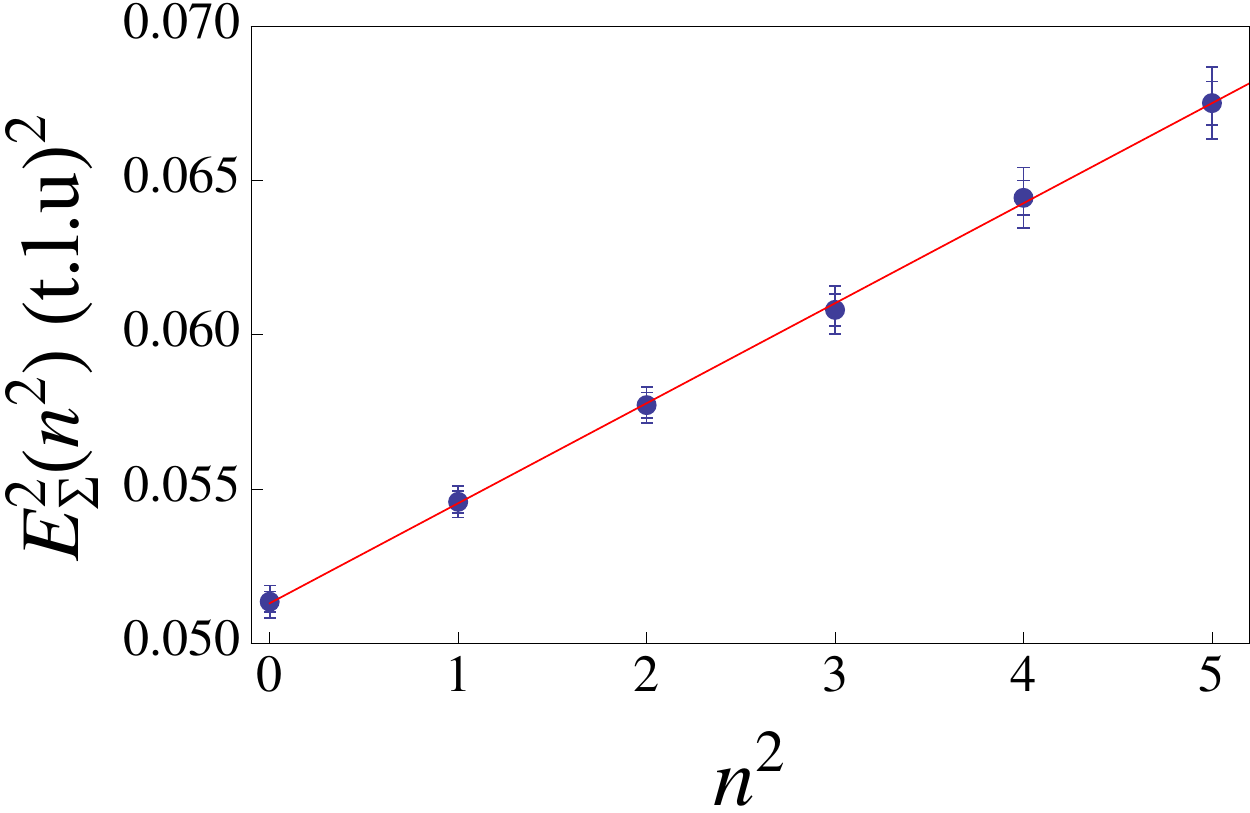}\ \ 
     \includegraphics[width=0.49\textwidth]{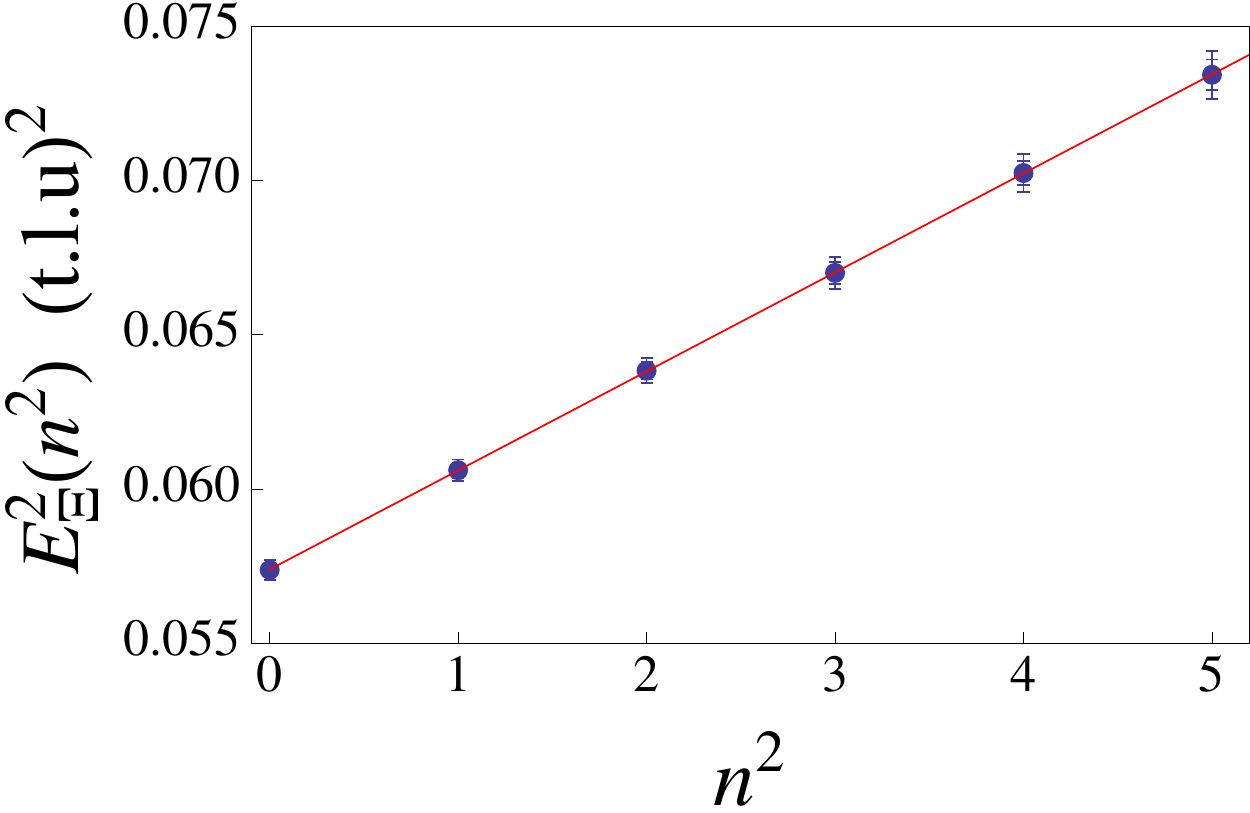}\ \ 
\caption{ The squared energy (in $({\rm t.l.u.})^2$) of the single baryon
  states as a function of
$n^2=|{\bf n}|^2$, related to the squared-momentum, 
$|{\bf p}|^2 = \left({2\pi\over L}\right)^2 |{\bf n}|^2$,
calculated with the $32^3\times 256$ ensemble.
The blue points are the results of the LQCD calculations with the inner (outer)
uncertainties being the statistical uncertainties (statistical and systematic
uncertainties combined in quadrature).
The red curves correspond to the best linear-fits.
}
  \label{fig:Dispersion}
\end{figure}

L\"uscher's method assumes that the single-hadron energy-momentum
relation is satisfied over the range of energies used in eq.~(\ref{eq:Luscher}).
In order to verify that the energy-momentum relation is satisfied,
single hadron correlation functions were formed with well-defined
lattice spatial momentum ${\bf p}={2\pi\over L} {\bf n}$ for $|{\bf
  n}|^2 \le 5$. 
Retaining the leading terms
in the energy-momentum relation, including the lattice anisotropy
$\xi$, the energy and mass of the hadron
(in temporal lattice units  (t.l.u)), and the momentum
in spatial lattice units (${\rm s.l.u}$) are related by
\begin{eqnarray}
\left(\,b_t\,E_H\left(|{\bf n}|^2\right)\right)^2 
& = &
(b_t\,M_H)^2 \ +\
\frac{1}{\xi^2}\,\left(\frac{2\,\pi\,b_s}{L}\right)^2\,|{\bf n}|^2 
\ .
\end{eqnarray}
The calculated single hadron energies (squared) are shown in
fig.~\ref{fig:Dispersion} as a function of $|{\bf n}|^2$,
along with the best linear fit. 
The extracted values of $\xi_H$ are given in
Table~\ref{tab:LQCDbaryondispersion}, and 
are seen to be consistent with each other within the uncertainties of the
calculation (the value for the nucleon is somewhat larger).
\begin{table}[!ht]
  \caption{The anisotropy parameter, $\xi_H$,  of each hadron from the
    $32^3\times 256$ ensemble.   The result for the $\pi$ is included for purposes of comparison.
  }
  \label{tab:LQCDbaryondispersion}
  \begin{ruledtabular}
    \begin{tabular}{c||ccccc}
          &  N &  $\Lambda$ &  $\Sigma$ & $\Xi$  & $\pi$ \\
      \hline
$\xi_H$ \ \ &  3.559(27)(08) & 3.465(31)(06) & 3.456(35)(07) & 3.4654(55)(14) &
3.466(13)(02) \\
    \end{tabular}
  \end{ruledtabular}
\end{table}
These values are used to
convert the two-hadron energies and energy differences from temporal lattice
units into spatial lattice units which  are then used in the L\"uscher
eigenvalue relation.

%%%%%%%%%%%%%%%%%%%%%%%%%%%%%%%%%%%%%%%%%%%%%%%%%%%%
\section{Two-Body Bound States}
\label{sec:BB}
\noindent
Of the baryon-baryon channels that we have explored  at this pion mass,
the states that have an energy lower than two
isolated baryons 
in both the $24^3\times 128$ and $32^3\times 256$ ensembles
and suggest the existence of bound states
are  the deuteron, the di-neutron, the H-dibaryon, and the $\Xi^-\Xi^-$.
While a negative energy shift can indicate either a scattering state with an
attractive interaction or a bound state, L\"uscher's eigenvalue relation
allows us to 
distinguish between the two possibilities.  
For a bound system in the large-volume limit, the
calculated value of the energy splitting (or binding momentum) gives rise to
$-i\cot\delta\rightarrow +1$.  
We now examine each of these channels.

%%%%%%%%%%%%%%%%%%%%%%%%%%%%%%%%%%%%%%%%%%%%%%%%%%%%
\subsection{The Deuteron}
\label{sec:deut}
\noindent
The deuteron is the simplest nucleus, comprised of a neutron and a proton.
At the physical light-quark masses
its binding energy is $B=2.224644(34)~{\rm MeV}$ which corresponds to a binding
momentum of $\kappa_0\sim 45.70~{\rm
  MeV}$ (using the isospin averaged nucleon mass of $M_N=938.92~{\rm MeV}$).
As it is a spin-1 system composed of two spin-${1\over 2}$ nucleons, its
wavefunction is an admixture of s-wave and d-wave, but 
at the physical quark masses
it is known to be predominantly
s-wave with only a small admixture of d-wave induced by the tensor 
($L=S=2$) interaction.

\begin{figure}[!ht]
  \centering
     \includegraphics[width=0.49\textwidth]{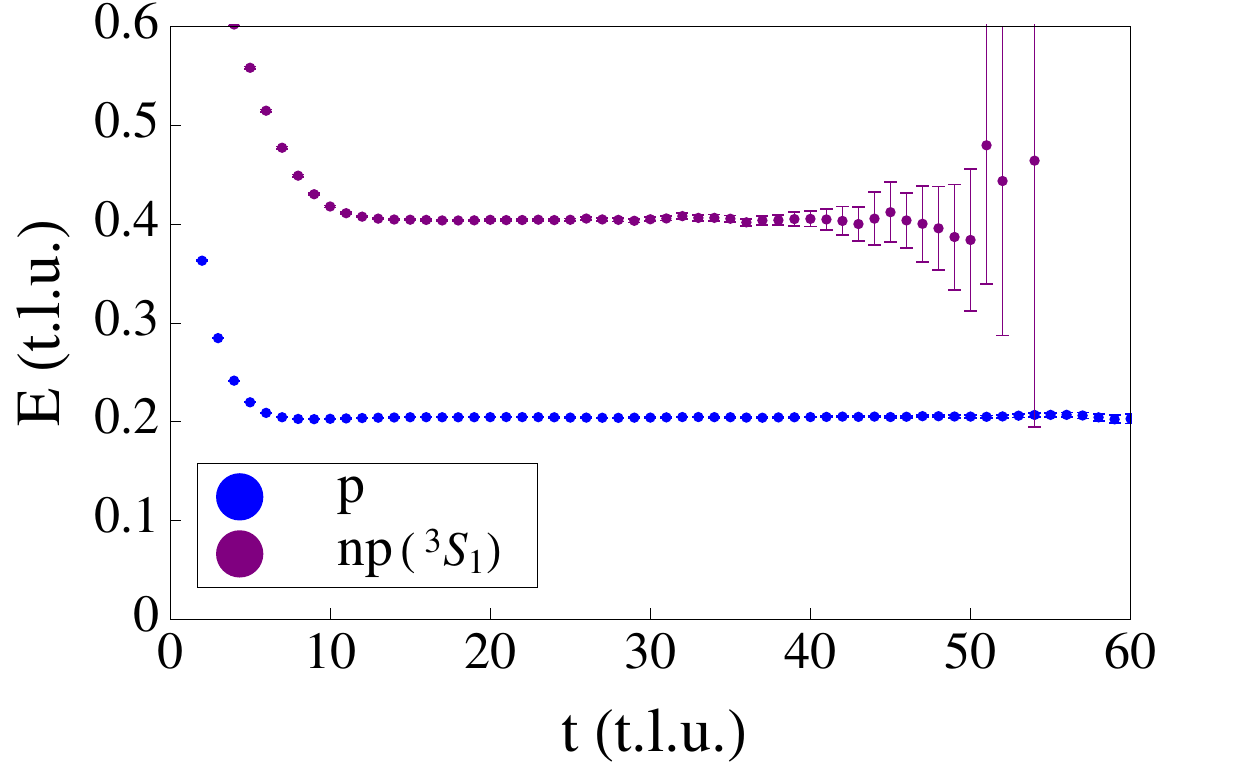}\ \  
     \includegraphics[width=0.49\textwidth]{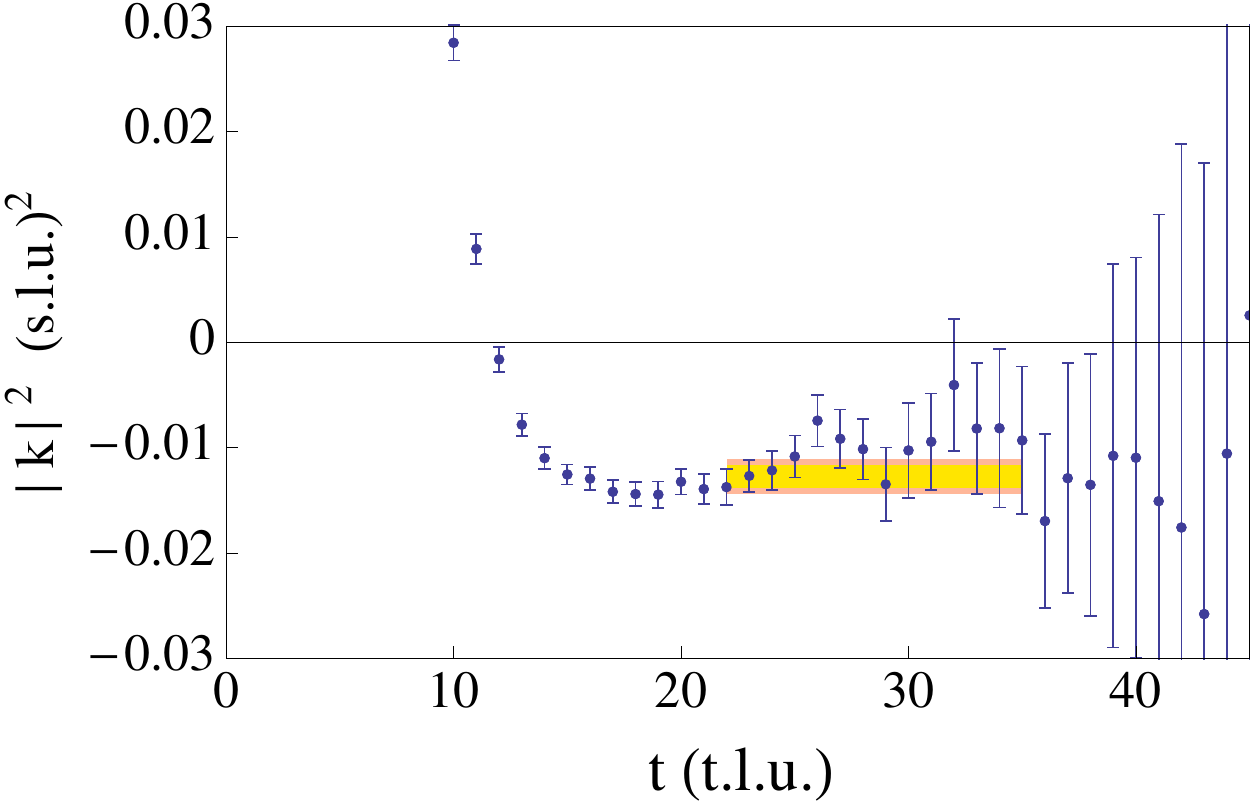}\ \  
\caption{The left panel shows an EMP of the nucleon and of the neutron-proton
  system in the $\siii-\diii$ coupled channels calculated with the $24^3\times
  128$ ensemble (in t.l.u.).  
The right panel shows the $|{\bf k}|^2$ (in $({\rm s.l.u.})^2$) of the
neutron-proton system calculated with this ensemble, along with the fits.
}
  \label{fig:Nemp24}
\end{figure}

\begin{figure}[!ht]
  \centering
     \includegraphics[width=0.49\textwidth]{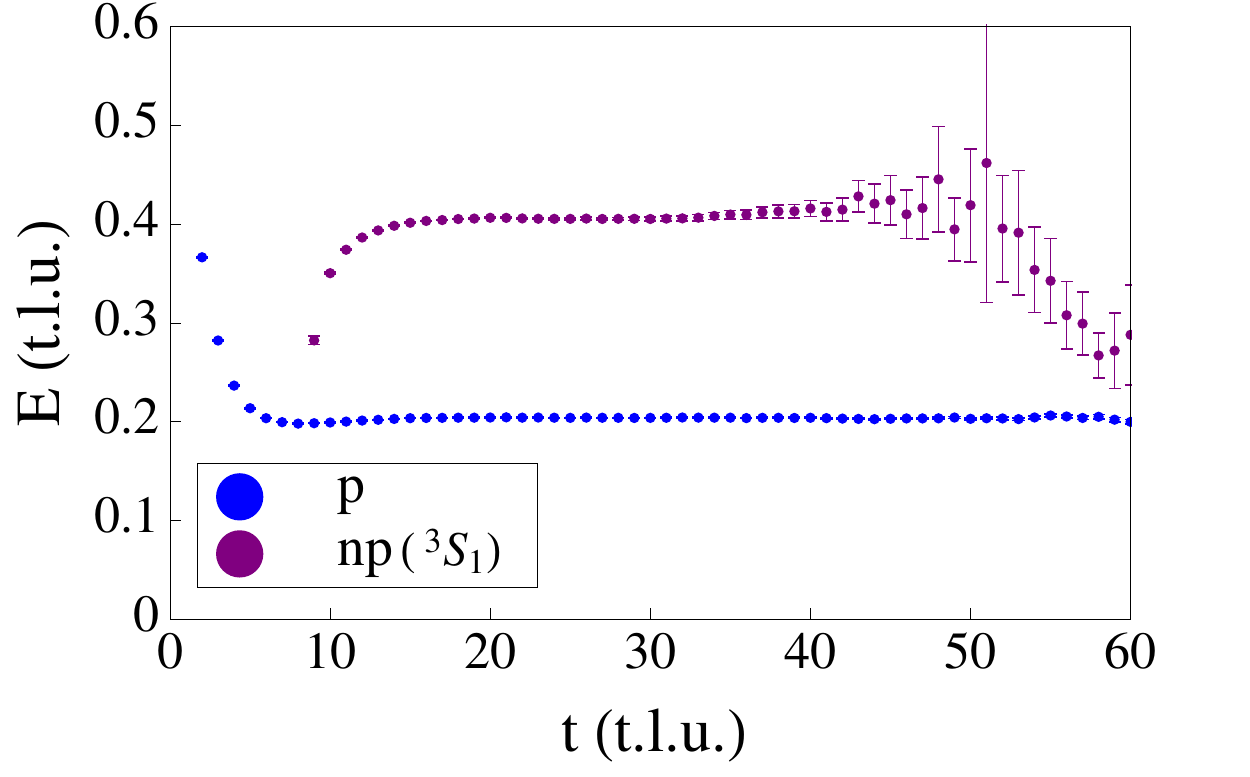}\ \  
     \includegraphics[width=0.49\textwidth]{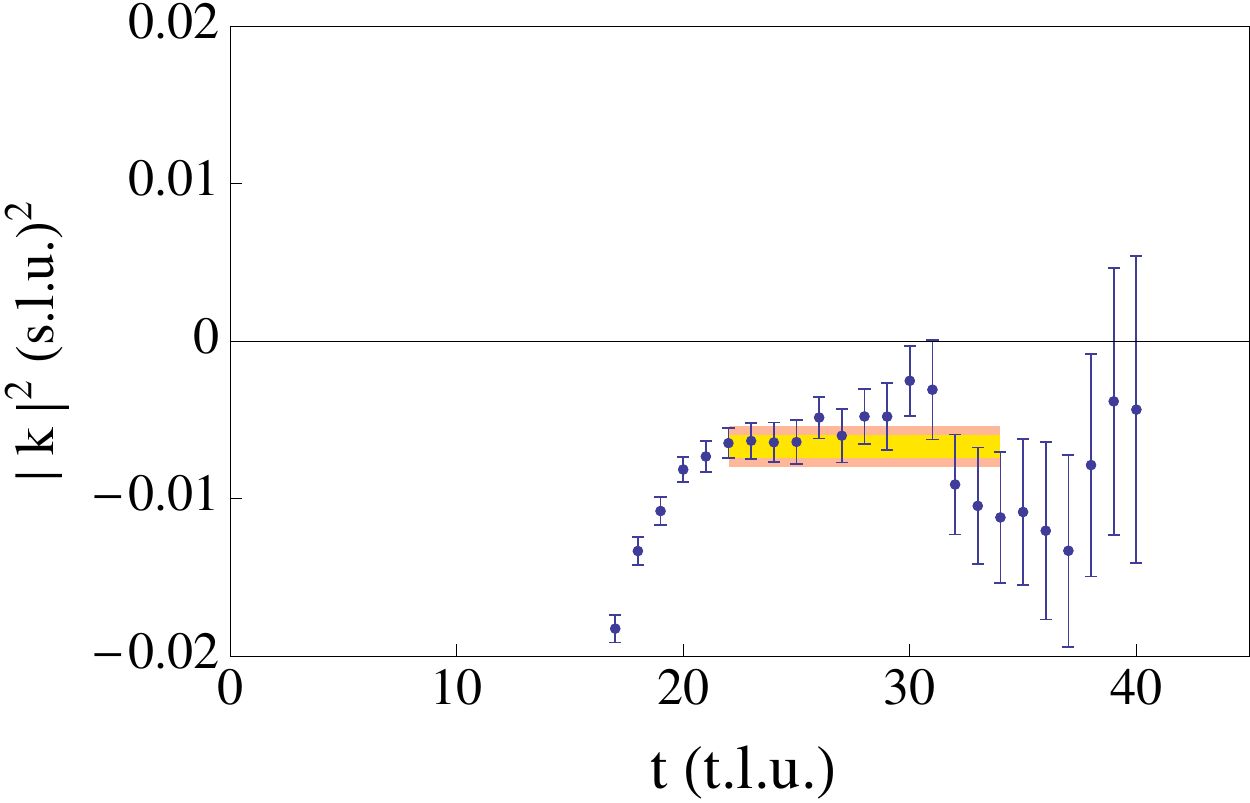}\ \  
\caption{The left panel shows an EMP of the nucleon and of the neutron-proton
  system in the $\siii-\diii$ coupled channels calculated with the 
$32^3\times 256$ ensemble (in t.l.u.).  
The right panel shows the $|{\bf k}|^2$ (in $({\rm s.l.u.})^2$) of the
neutron-proton system calculated with this ensemble, along with the fits.
}
  \label{fig:Nemp32}
\end{figure}

\begin{figure}[!ht]
  \centering
     \includegraphics[width=0.7\textwidth]{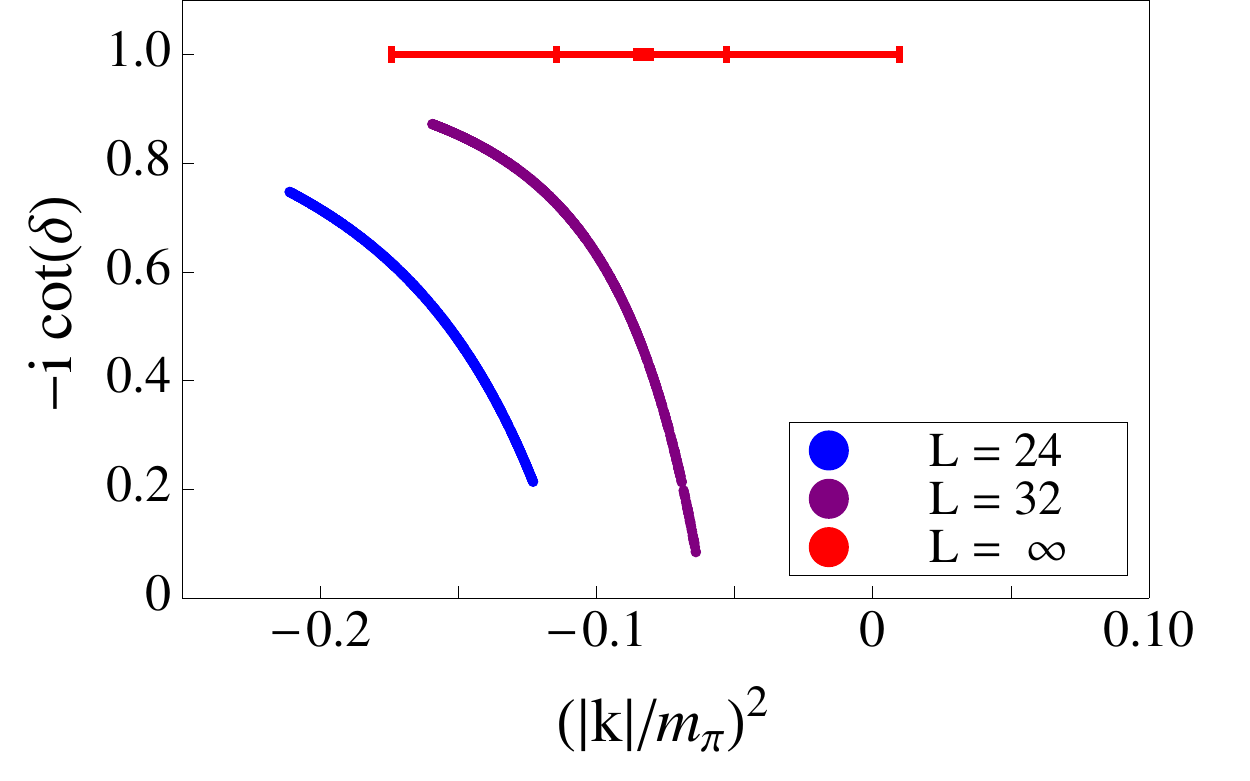}
\caption{Results of the Lattice QCD calculations of $-i\cot\delta$ 
versus $|{\bf k}|^2/m_\pi^2$ 
in the deuteron channel
obtained using eq.~(\protect\ref{eq:Luscher}),
along with the infinite-volume extrapolation using
eq.~(\protect\ref{eq:pcotkappa}).
The inner uncertainty of the infinite-volume extrapolation is statistical,
while the outer corresponds to the statistical and systematic uncertainties
combined in quadrature. 
}
  \label{fig:NNextrap}
\end{figure}

The EMPs associated with the nucleon and the
neutron-proton system in the $\siii-\diii$ channel are shown in the left panels
of fig.~\ref{fig:Nemp24} and fig.~\ref{fig:Nemp32} for the two ensembles. 
The correlation functions that give rise to these EMPs are linear combinations
of correlation functions generated using eq.~(\ref{eq:correlators}) but with
different smearings of the sink operator(s).
The combinations of correlation functions have been chosen  to maximize the
extent of the ground-state 
plateaus~\footnote{The EMPs result from a matrix-Prony
  analysis~\cite{Beane:2009kya} 
of multiple correlation functions.  
In determining the binding energies, 
multi-exponential fitting and generalized pencil of function (GPoF) 
methods~\cite{Aubin:2010jc,ko} are used
in
addition to Matrix-Prony and provides consistent results in each case.
}.
Extended plateaus are observed in both the one and two nucleon correlation
functions.
The right panels of fig.~\ref{fig:Nemp24} and fig.~\ref{fig:Nemp32} show the
binding momentum of each particle
in the CoM obtained by taking ratios of the two-baryon and single-baryon
correlation functions.
The deuteron binding energies in each volume calculated with LQCD are
\begin{eqnarray}
B_d^{(L=24)} & = & 
22.3\pm 2.3\pm 5.4~{\rm MeV}
\ \ ,\ \ 
B_d^{(L=32)}\ =\ 14.9\pm 2.3\pm 5.8~{\rm MeV}
\ \ \ .
\label{eq:deutbindingLQCD}
\end{eqnarray}
The known finite-volume dependence of loosely bound systems, given in
eq.~(\ref{eq:pcotkappa}), and the perturbative relations that follow, 
allow for an extrapolation of the results in eq.~(\ref{eq:deutbindingLQCD})
to the infinite-volume limit,
as shown in fig.~\ref{fig:NNextrap}, giving
\begin{eqnarray}
B_d^{(L=\infty)} & = & 
11\pm 5\pm 12~{\rm MeV}
\ \ \ 
\label{eq:deutbindingLQCDextrap}
\end{eqnarray}
where the 
first uncertainty is 
statistical and the second is 
systematic,
accounting for fitting, anisotropy, lattice spacing  and the infinite volume extrapolation.
Despite having statistically significant binding energies in the two lattice
volumes, the exponential extrapolation to the
infinite volume limit produces a deuteron binding energy with significance at
$\sim 1\sigma$. 
If the uncertainties of both LQCD calculations were reduced by a factor of two,
the significance
of the extrapolated binding energy would increase to $\sim 3\sigma$ if the
central values remained unchanged.
From the curvature of the results of the LQCD calculations in
fig.~\ref{fig:NNextrap}, it is clear the both of these volumes 
significantly modify 
the deuteron at this pion mass.  Calculations in somewhat
larger volumes, or of moving systems~\cite{Davoudi:2011md}, 
would significantly  reduce the uncertainty introduced by the volume
extrapolation.

\begin{figure}[!ht]
  \centering
     \includegraphics[width=0.7\textwidth]{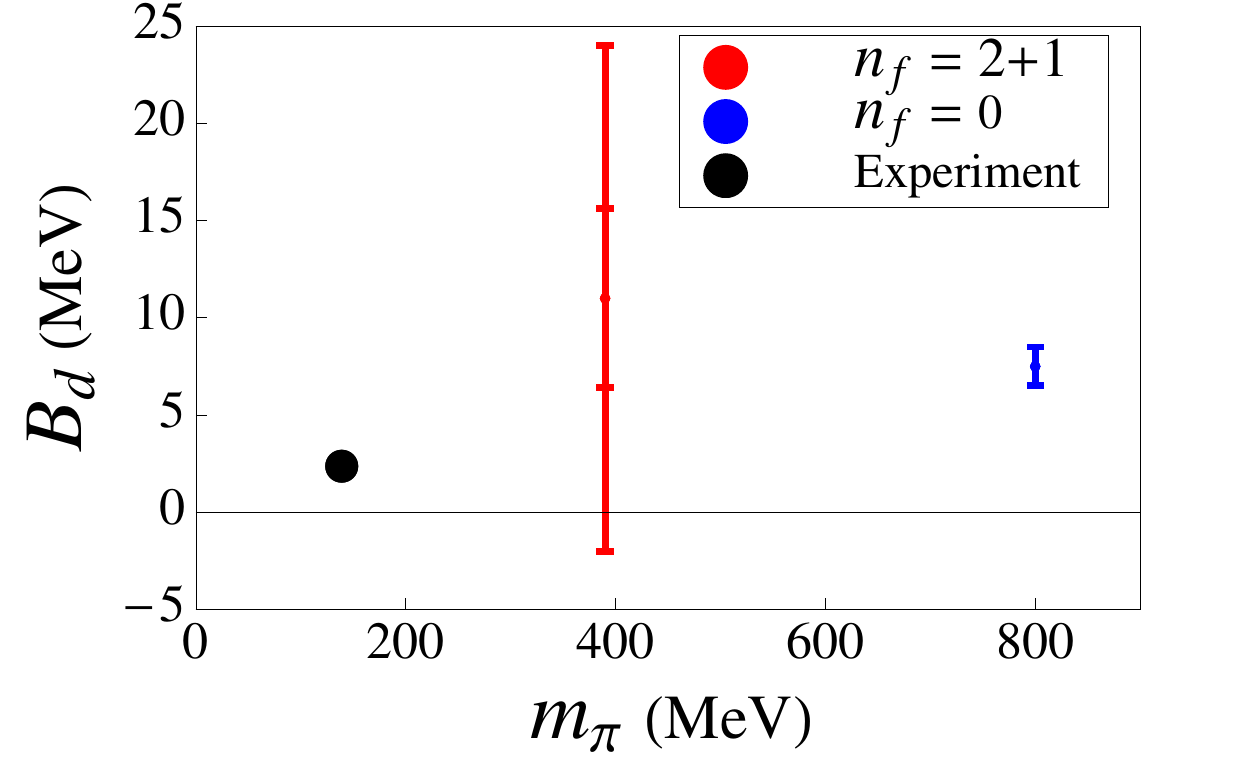}
\caption{The deuteron binding energy as a function of the pion mass.
The black circle denotes the experimental value.
The blue point and uncertainty results from the quenched
calculations of Ref.~\protect\cite{Yamazaki:2011nd}, while the red point and
uncertainty 
(the inner is statistical and the outer is statistical and systematic
combined in quadrature)
is our present $n_f=2+1$ result.
}
  \label{fig:DeutALL}
\end{figure}
Our $n_f=2+1$ result and the recent quenched ($n_f=0$) result of
Ref.~\cite{Yamazaki:2011nd} are shown in fig.~\ref{fig:DeutALL}, along with the
physical deuteron binding energy.
Clearly, the large uncertainty of our present result does not provide much 
constraint on the dependence of the deuteron binding energy as a
function of the light-quark masses, other than to demonstrate that the deuteron
is likely bound at $m_\pi\sim 390~{\rm MeV}$, qualitatively consistent with the
quenched result at $m_\pi\sim 800~{\rm MeV}$~\cite{Yamazaki:2011nd}.

A number of groups have attempted 
to 
determine how the deuteron binding energy (and the binding of other nuclei) 
varies as a function of the
light-quark masses using EFT~\cite{Beane:2002vs,Epelbaum:2002gb,Beane:2002xf,Chen:2010yt}
and hadronic models~\cite{Flambaum:2007mj}.
Such a variation impacts the constraints that can be placed on possible time-variations
of the fundamental constants of nature from the abundance of elements produced
in Big Bang Nucleosynthesis (BBN) (see
Refs.~\cite{Bedaque:2010hr,Cheoun:2011yn} for recent constraints from
BBN).
With the exception of the analysis of Ref.~\cite{Chen:2010yt}, 
both of the EFT analyses, which use naive dimensional analysis (NDA) to constrain
the quark-mass dependent dimension-six operators that contribute at next-to-leading
order (NLO) in the chiral
expansion, and the hadronic models 
of Ref.~\cite{Flambaum:2007mj},
suggest that
the deuteron becomes less bound as the quarks become heavier near their
physical values. 
The present LQCD calculation at a pion mass of $m_\pi\sim 390~{\rm MeV}$ is
somewhat beyond the range of applicability of the EFT analyses and so
cannot be directly translated into constraints on the coefficients of local
operators with confidence.  
Further, the uncertainty in our calculation is too large to be useful in a
quantitative way.  
Nevertheless, our result conflicts with the trend suggested in most of the EFT and
model analyses, and further studies are necessary to resolve this issue.

%%%%%%%%%%%%%%%%%%%%%%%%%%%%%%%%%%%%%%%%%%%%%%%%%%%%
\subsection{The Di-Neutron}
\label{sec:NN1s0}
\noindent
In nature, the di-neutron ($nn$ $\si$)
is very nearly bound.  The unnaturally large scattering
lengths in the $\si$-channel indicate that a very small increase in the
strength of the interactions between neutrons would bind them into an
electrically neutral nucleus.  If the binding was deep enough, it would have
profound effects on nucleosynthesis.
Analyses with NNEFT allow for the possibility of both bound and unbound
di-neutrons for light-quark masses larger than those of nature, while
indicating an unbound di-neutron for lighter quark 
masses~\cite{Beane:2002vs,Epelbaum:2002gb,Beane:2002xf}.
In contrast, a model-dependent calculation indicates that the di-neutron
remains unbound for all light-quark masses~\cite{Flambaum:2007mj}.

\begin{figure}[!ht]
  \centering
     \includegraphics[width=0.49\textwidth]{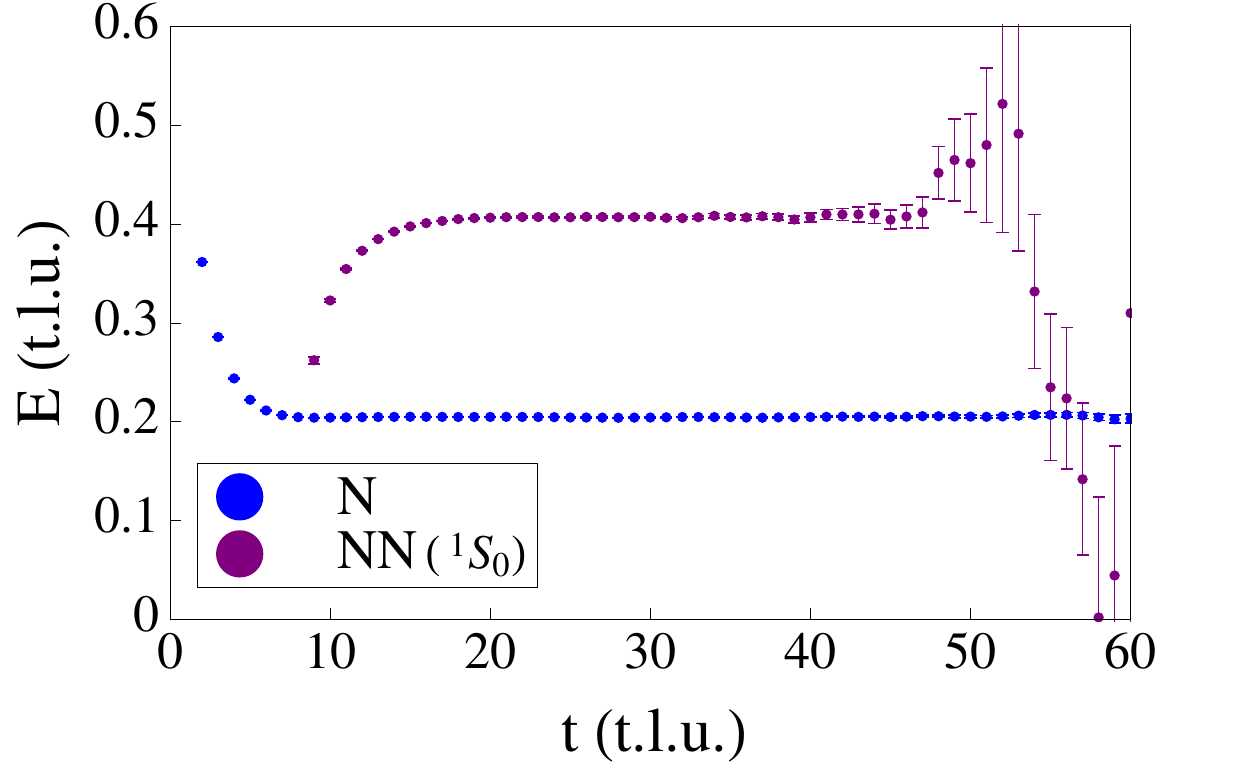}\ \  
     \includegraphics[width=0.49\textwidth]{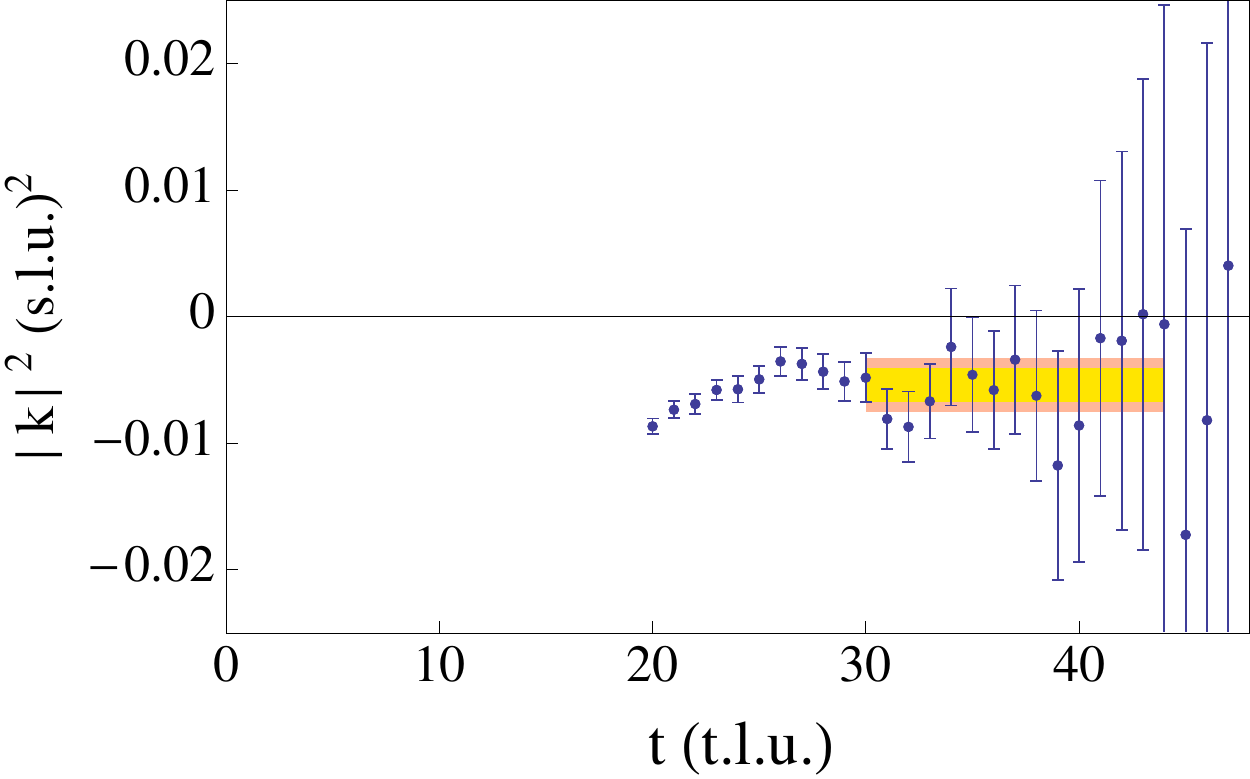}\ \  
\caption{The left panel shows an EMP of the neutron  and of the neutron-neutron
  system calculated with the $24^3\times
  128$ ensemble (in t.l.u.).  
The right panel shows the $|{\bf k}|^2$ (in $({\rm s.l.u.})^2$) of the
neutron-neutron  system calculated with this ensemble, along with the fits.
}
  \label{fig:NN1s0emp24}
\end{figure}

\begin{figure}[!ht]
  \centering
     \includegraphics[width=0.49\textwidth]{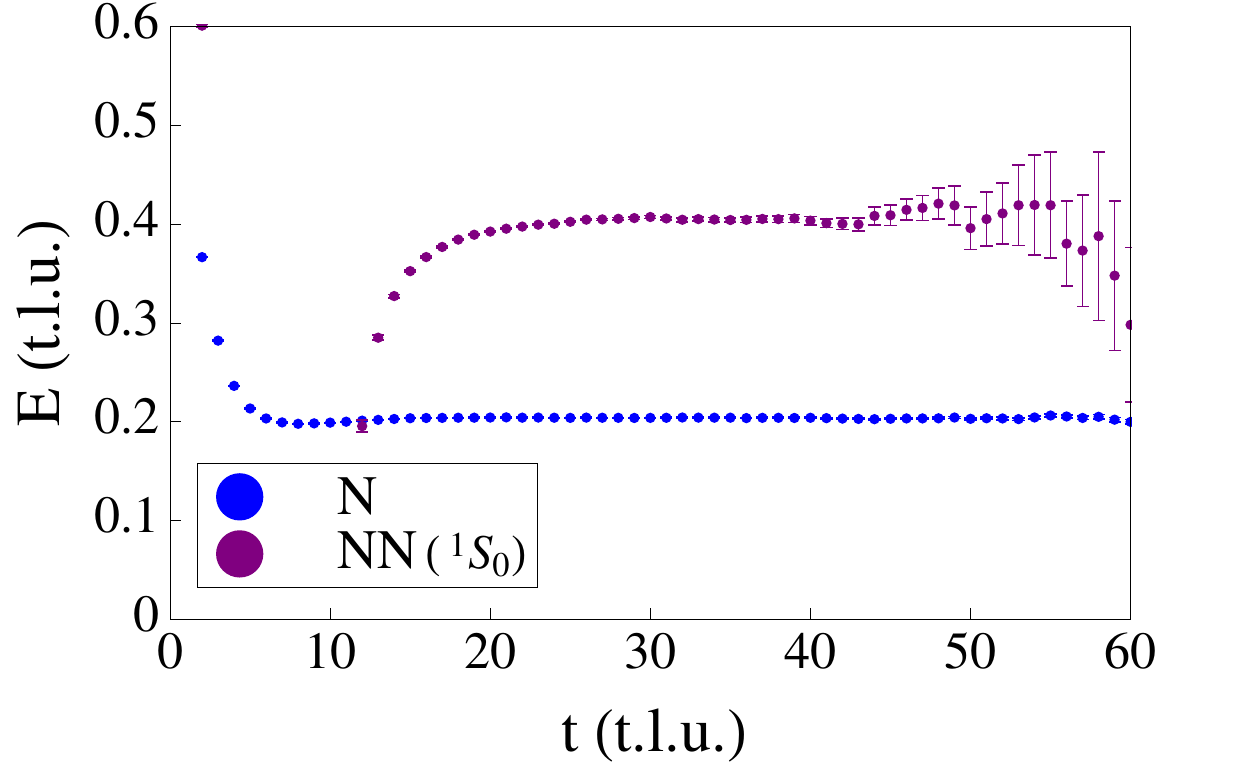}\ \  
     \includegraphics[width=0.49\textwidth]{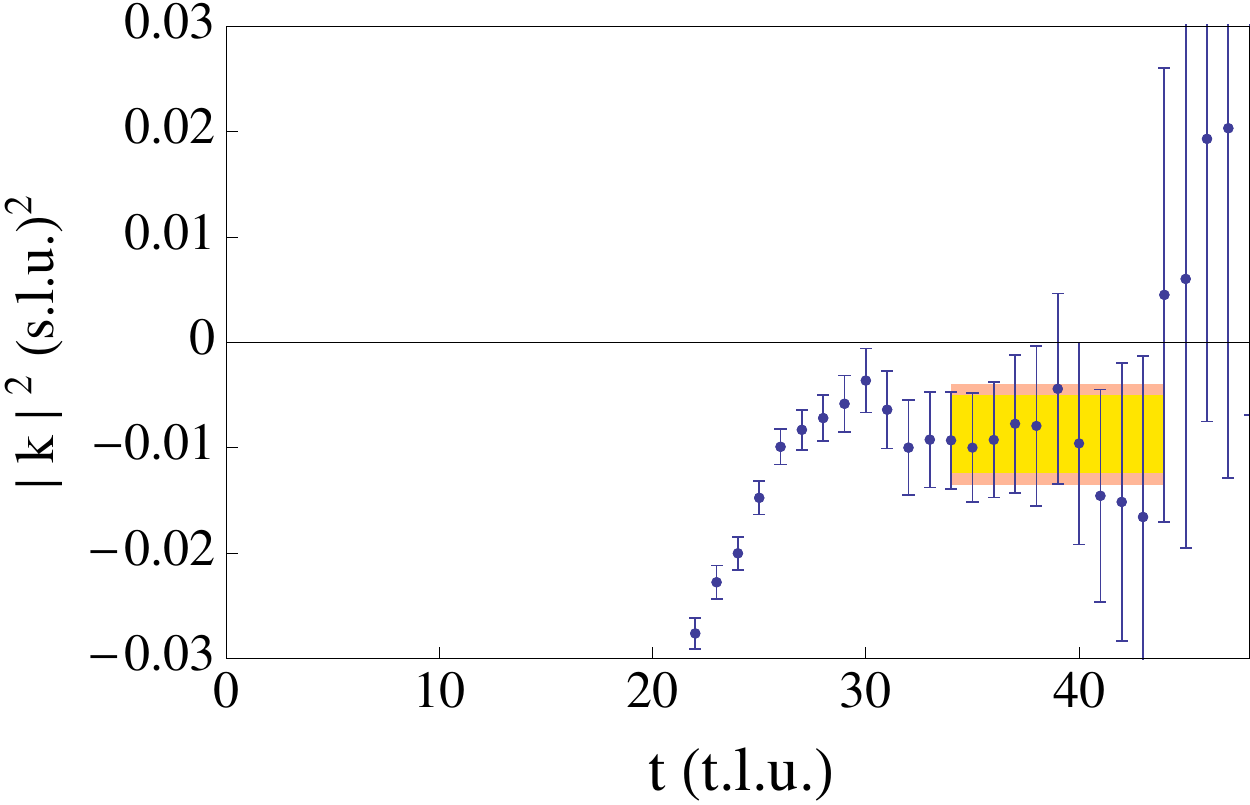}\ \  
\caption{The left panel shows an EMP of the neutron  and of the neutron-neutron
  system calculated with the $32^3\times 256$ ensemble (in t.l.u.).  
The right panel shows the $|{\bf k}|^2$ (in $({\rm s.l.u.})^2$) of the
neutron-neutron  system calculated with this ensemble, along with the fits.
}
  \label{fig:NN1s0emp32}
\end{figure}

\begin{figure}[!ht]
  \centering
     \includegraphics[width=0.7\textwidth]{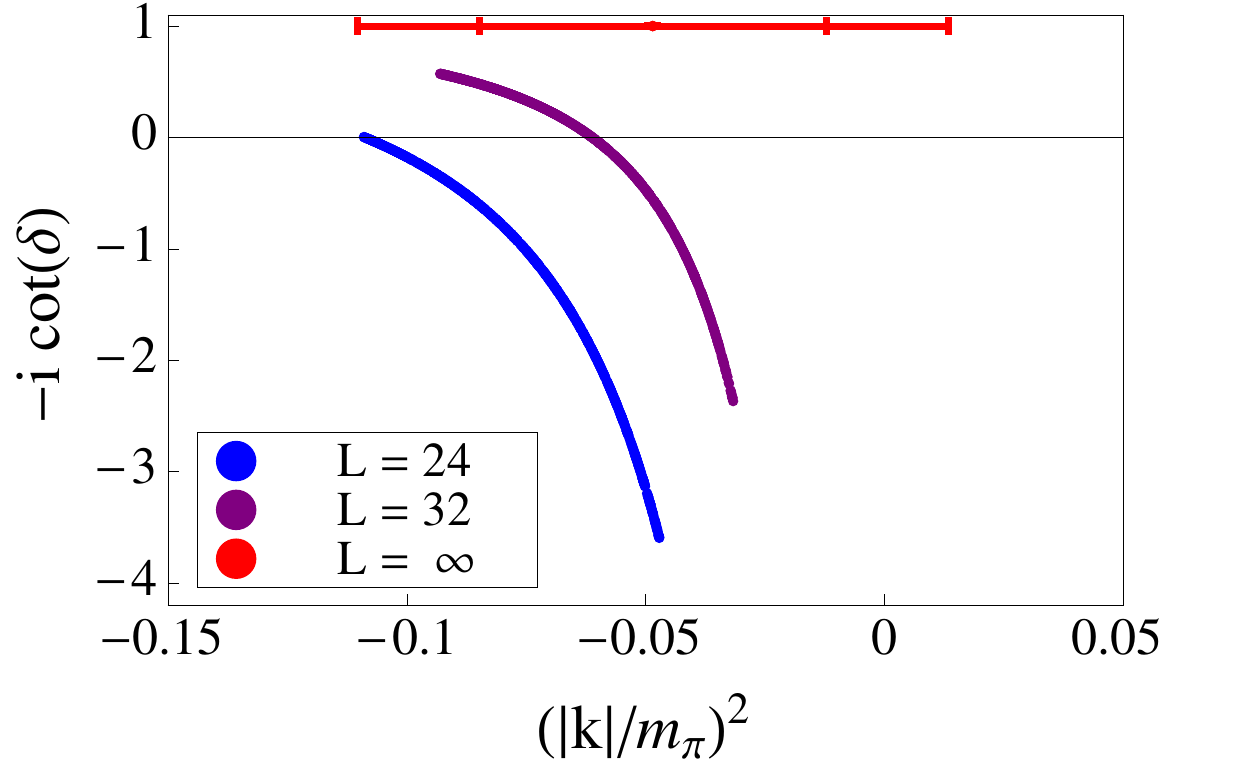}
\caption{The results of the Lattice QCD calculations of $-i\cot\delta$ 
versus $|{\bf k}|^2/m_\pi^2$ 
in the di-neutron channel
obtained using eq.~(\protect\ref{eq:Luscher}),
along with the infinite-volume extrapolation using eq.~(\protect\ref{eq:pcotkappa}).
The inner uncertainty of the infinite-volume extrapolation is statistical,
while the outer corresponds to the statistical and systematic uncertainties
combined in quadrature. 
}
  \label{fig:NN1s0extrap}
\end{figure}

The EMPs associated with the nucleon and the
di-neutron system are shown in the left panels
of fig.~\ref{fig:NN1s0emp24} and fig.~\ref{fig:NN1s0emp32}. 
The di-neutron  binding energies extracted from the 
LQCD calculations are
\begin{eqnarray}
B_{nn}^{(L=24)} & = & 
10.4\pm 2.6\pm 3.1~{\rm MeV}
\ \ ,\ \ 
B_{nn}^{(L=32)}\ =\ 8.3\pm 2.2\pm 3.3~{\rm MeV}
\ \ \ .
\label{eq:nnbindingLQCD}
\end{eqnarray}
The volume extrapolation of the  results in  eq.~(\ref{eq:nnbindingLQCD})
is shown in fig.~\ref{fig:NN1s0extrap}, and results in 
an extrapolated di-neutron binding energy of
\begin{eqnarray}
B_{nn}^{(L=\infty)} & = & 
7.1\pm 5.2\pm 7.3~{\rm MeV}
\ \ \ 
\label{eq:nnbindingLQCDextrap}
\end{eqnarray}
where the first uncertainty is statistical and the second is systematic.
This result is suggestive of a bound di-neutron at this pion mass, 
but at the present level of precision an unbound system is also possible.

\begin{figure}[!ht]
  \centering
     \includegraphics[width=0.7\textwidth]{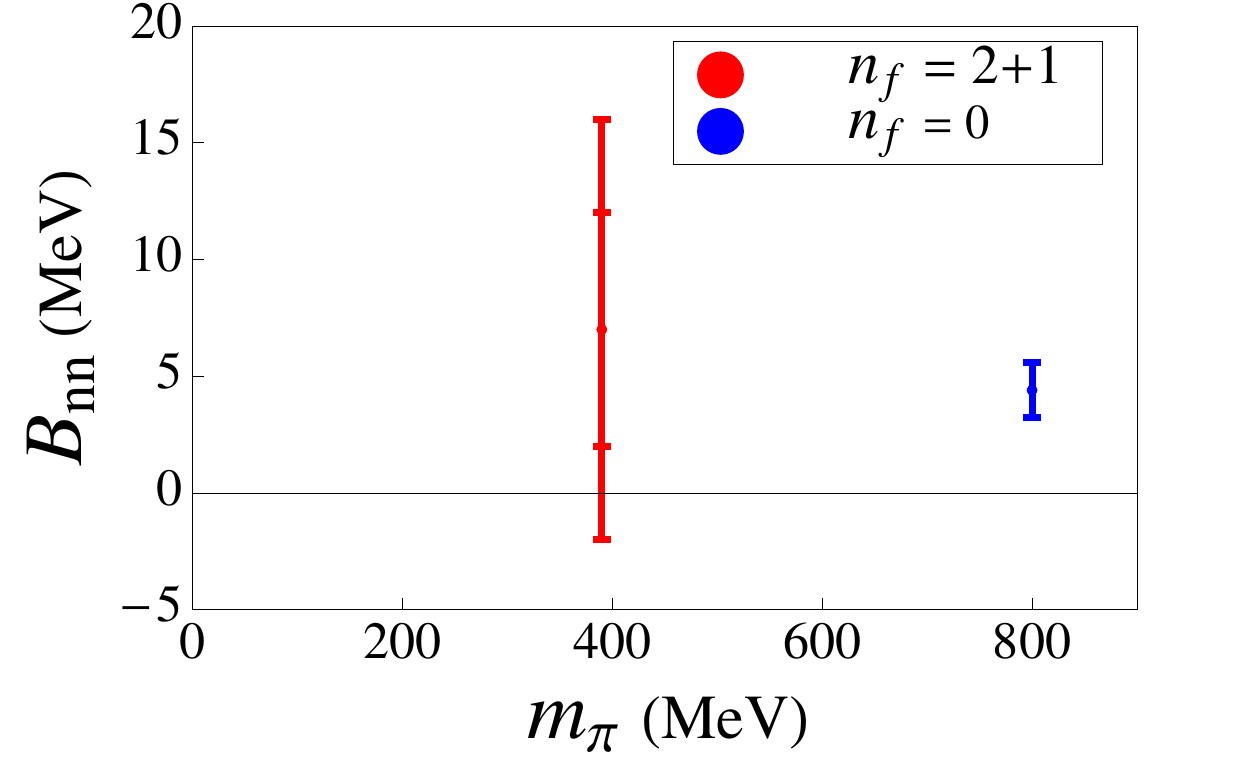}
\caption{The di-neutron binding energy as a function of the pion mass.
The blue point and uncertainty results from the quenched
calculation of Ref.~\protect\cite{Yamazaki:2011nd}, while the red point and
uncertainty 
(the inner is statistical and the outer is statistical and systematic
combined in quadrature)
is our present $n_f=2+1$ result.
}
  \label{fig:nnALL}
\end{figure}
Our $n_f=2+1$ result and the recent quenched ($n_f=0$) result of
Ref.~\cite{Yamazaki:2011nd} are shown in fig.~\ref{fig:nnALL}.
Clearly, the large uncertainty of our present result does not provide a
significant constraint on the binding of the di-neutron
as a
function of the light-quark masses.
However, the LQCD results suggest that the di-neutron is bound
at quark masses greater than those of nature.  
This has implication for future LQCD calculations as there are likely
light-quark masses for which the di-neutron unbinds, and hence the scattering
length becomes infinitely large.  This implies that, at some point in the future,
LQCD may be able to explore strongly interacting 
systems of fermions near the unitary limit.
However, if the deuteron remains bound
at heavier quark masses, as suggested by the current work, 
it may not be possible to tune the light-quark
masses (including isospin breaking)
to produce infinite scattering lengths in the $\siii-\diii$ and $\si$
channels simultaneously and hence eliminating the possibility of the triton
having an infinite number of bound states for such a specific choice of
light-quark masses 
(unless the deuteron 
is also unbound for an intermediate range
of quark 
masses)~\footnote{Such bound states would be the  manifestation of an infrared
renormalization group limit cycle in QCD, as conjectured by Braaten and Hammer~\cite{Braaten:2003eu}.}.

%%%%%%%%%%%%%%%%%%%%%%%%%%%%%%%%%%%%%%%%%%%%%%%%%%%%
\subsection{The H-Dibaryon}
\label{sec:H}
\noindent
The prediction of a relatively deeply bound system with the quantum
numbers of $\Lambda\Lambda$ (called the H-dibaryon) by
Jaffe~\cite{Jaffe:1976yi} in the late 1970s, based upon a bag-model
calculation, started a vigorous search for such a system, both
experimentally and also with alternate theoretical tools.
As all six quarks, $uuddss$, can be in an s-wave and satisfy the
Pauli principle, such a channel may support a state that is more
deeply bound than in channels with different flavor quantum numbers.
Reviews of experimental constraints on, and phenomenological models of, the
H-dibaryon can be found in Refs.~\cite{Bashinsky:1997qv,Yamamoto:2000wf,Sakai:1999qm,Mulders:1982da}.
While experimental studies of doubly-strange ($s=-2$)
hypernuclei restrict
the H-dibaryon to be unbound or to have a small binding energy,
the most recent constraints on the existence of the H-dibaryon come
from heavy-ion collisions at RHIC~\cite{Trattner:2006jn}, effectively eliminating the
possibility of a loosely-bound H-dibaryon at the physical light-quark
masses. However, the analysis that led to these constraints was
model-dependent, in particular in the production mechanism, and may simply 
not be reliable.
Recent experiments at KEK indicate that a near threshold resonance may exist
in   this channel~\cite{Yoon:2007aq}.

A number of quenched LQCD
calculations~\cite{Mackenzie:1985vv,Iwasaki:1987db,Pochinsky:1998zi,Wetzorke:1999rt,Wetzorke:2002mx,Luo:2007zzb}
have previously searched for the H-dibaryon,  but without 
success.
Recently, we and the HALQCD collaboration have reported results that
show that the H-dibaryon is bound for a range of light-quark masses
that are larger than those found in nature~\cite{Beane:2010hg,Inoue:2010es}. 
At present, neither of these calculations are extrapolated to the continuum, with both
calculations being performed at a spatial lattice spacing of $b_s\sim 0.12~{\rm
  fm}$.
Chiral extrapolations in the light-quark masses
of these two LQCD calculations were performed in
Refs.~\cite{Beane:2011xf,Shanahan:2011su} to make first QCD predictions for the
binding energy of the H-dibaryon at the physical light-quark masses.~\footnote{
These extrapolations are significantly less reliable (rigorous) than the chiral
extrapolation of simple quantities (such as hadron masses) calculated with LQCD.
While for a deeply bound H-dibaryon with a  radius that is much smaller than the
inverse pion mass it is possible to arrive at a chiral EFT construction 
with
which to calculate the light-quark mass dependence of H-dibaryon mass in perturbation
theory, the same construction would not be valid when the radius becomes comparable to
or larger than $1/m_\pi$.  
A weakly  bound state can only be generated nonperturbatively, and
consequently the quark-mass dependence of the binding energy is  nontrivial, as
is clear from the analyses in the two-nucleon sector, 
e.g. Refs.~\cite{Beane:2002vs,Epelbaum:2002gb,Beane:2002xf,Mondejar:2006yu}.
As a result, the assumption of compactness of the state made in Ref.~\cite{Shanahan:2011su}
is difficult to justify over a significant range of predicted binding energies.
Further, the simple polynomial extrapolations in Ref.~\cite{Beane:2011xf} are
meant to provide estimates alone and cannot be used to reliably quantify
extrapolation uncertainties.
}

\begin{figure}[!ht]
  \centering
     \includegraphics[width=0.49\textwidth]{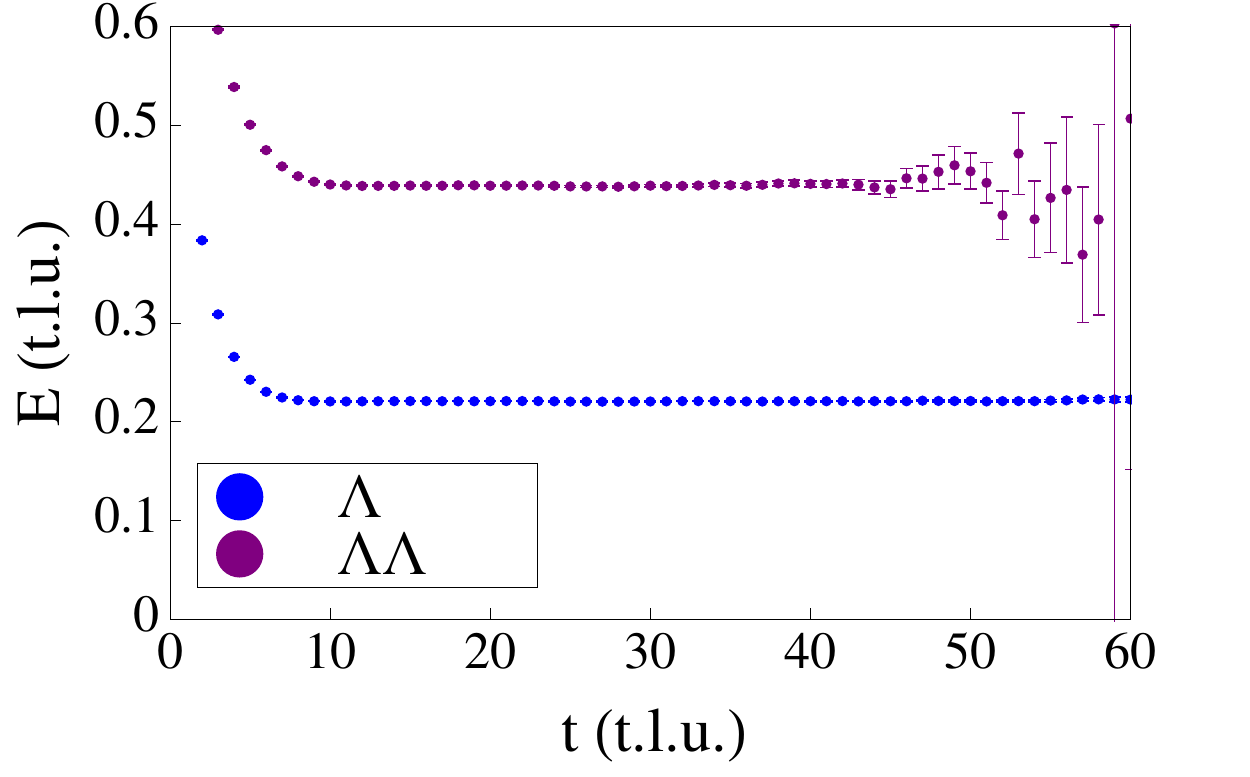}\ \  
     \includegraphics[width=0.49\textwidth]{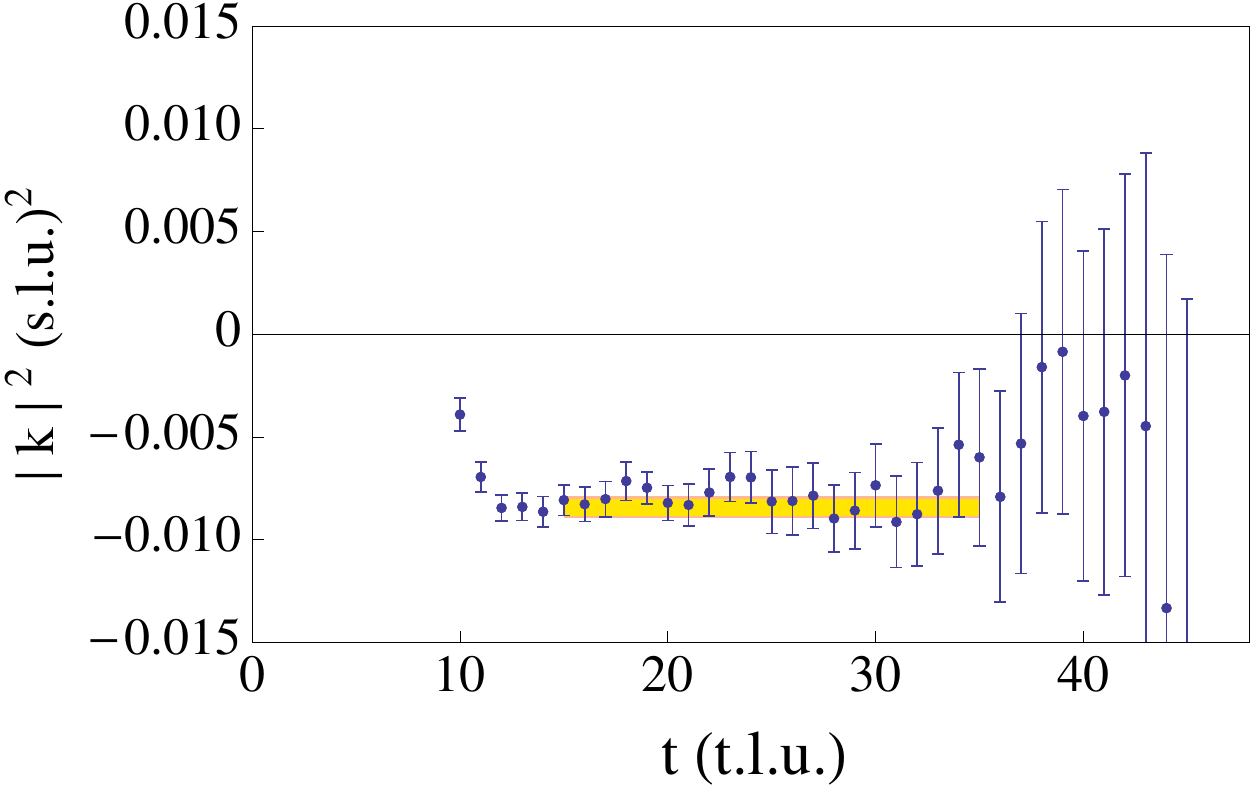}\ \  
\caption{The left panel shows an EMP of the $\Lambda$ and of the 
lowest state in the 
$\Lambda\Lambda$-$\Xi N$-$\Sigma\Sigma$
  system calculated with the $24^3\times
  128$ ensemble (in t.l.u.).  
The right panel shows the $|{\bf k}|^2$ (in $({\rm s.l.u.})^2$) of the
$\Lambda\Lambda$-$\Xi N$-$\Sigma\Sigma$ system calculated with this ensemble, along with the fits.
}
  \label{fig:Lamemp24}
\end{figure}

\begin{figure}[!ht]
  \centering
     \includegraphics[width=0.49\textwidth]{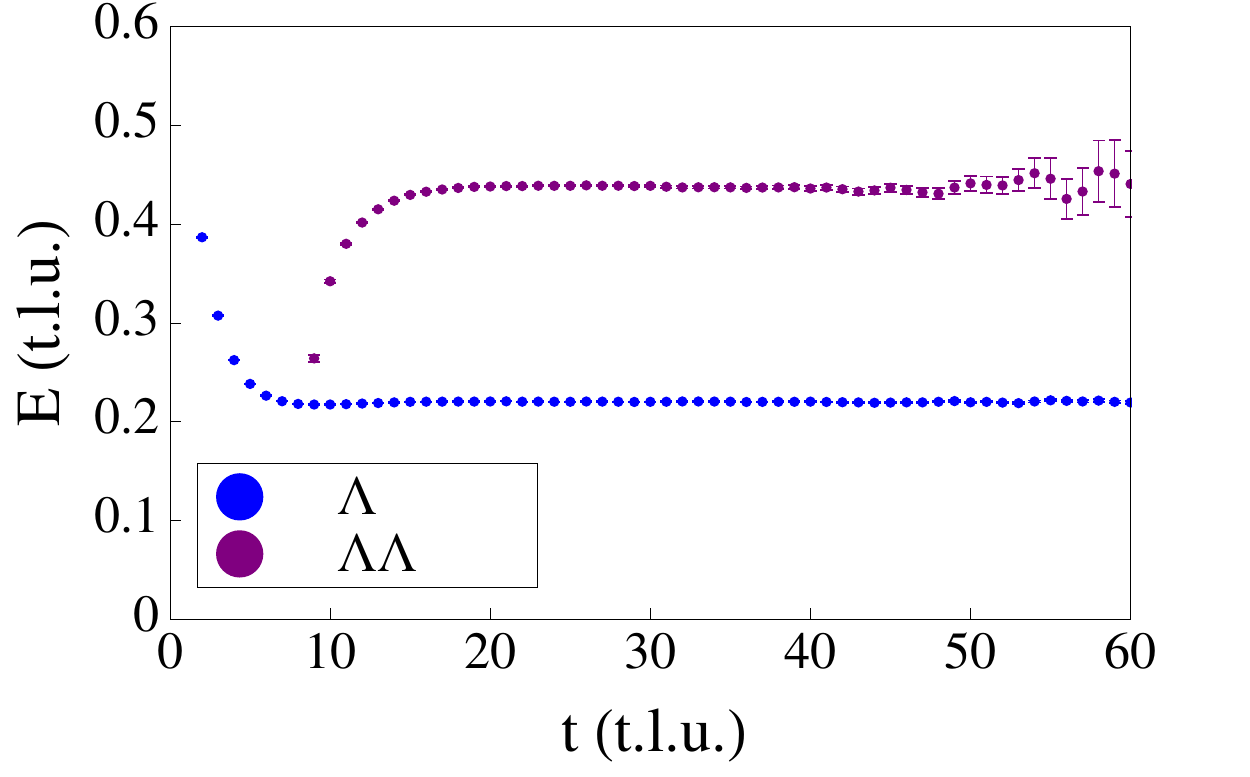}\ \  
     \includegraphics[width=0.49\textwidth]{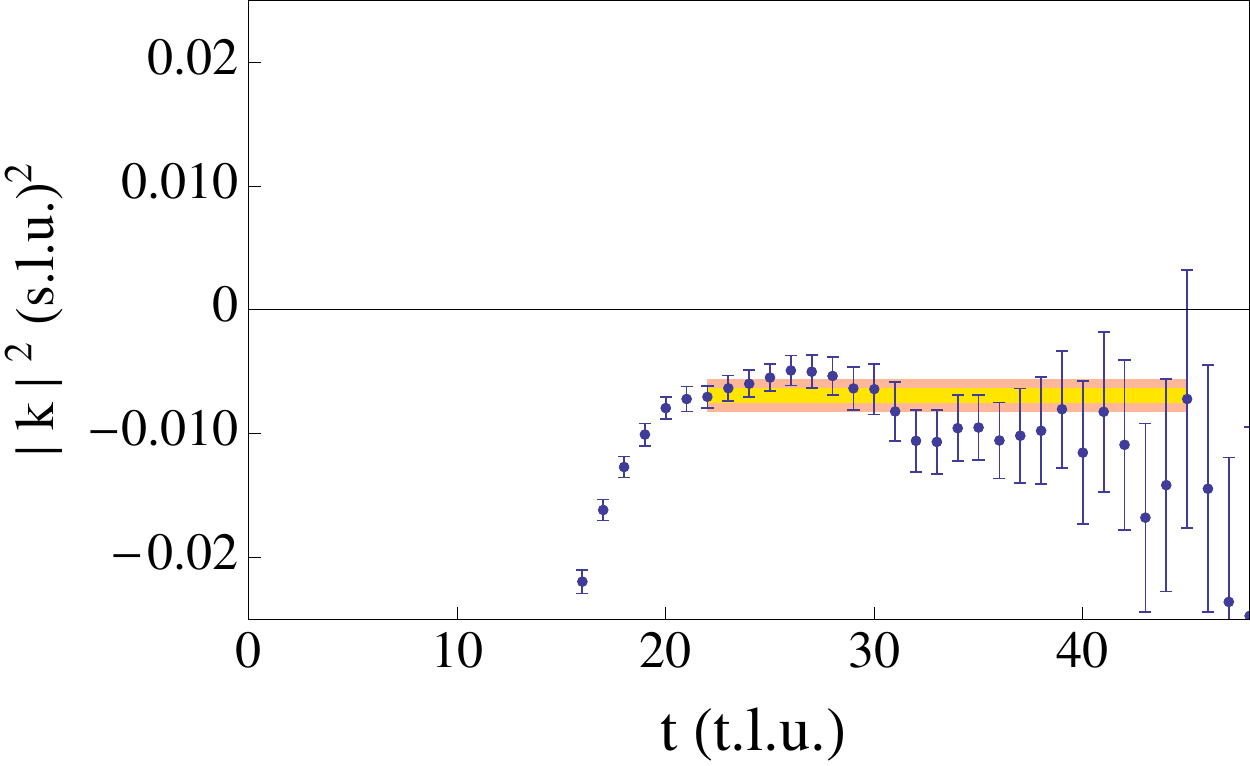}\ \  
\caption{The left panel shows an EMP of the $\Lambda$ and of the lowest state
  in the 
$\Lambda\Lambda$-$\Xi N$-$\Sigma\Sigma$
  system calculated with the 
$32^3\times 256$ ensemble (in t.l.u.).  
The right panel shows the $|{\bf k}|^2$ (in $({\rm s.l.u.})^2$) of the
$\Lambda\Lambda$-$\Xi N$-$\Sigma\Sigma$ system calculated with this ensemble, along with the fits.
}
  \label{fig:Lamemp32}
\end{figure}

\begin{figure}[!ht]
  \centering
     \includegraphics[width=0.7\textwidth]{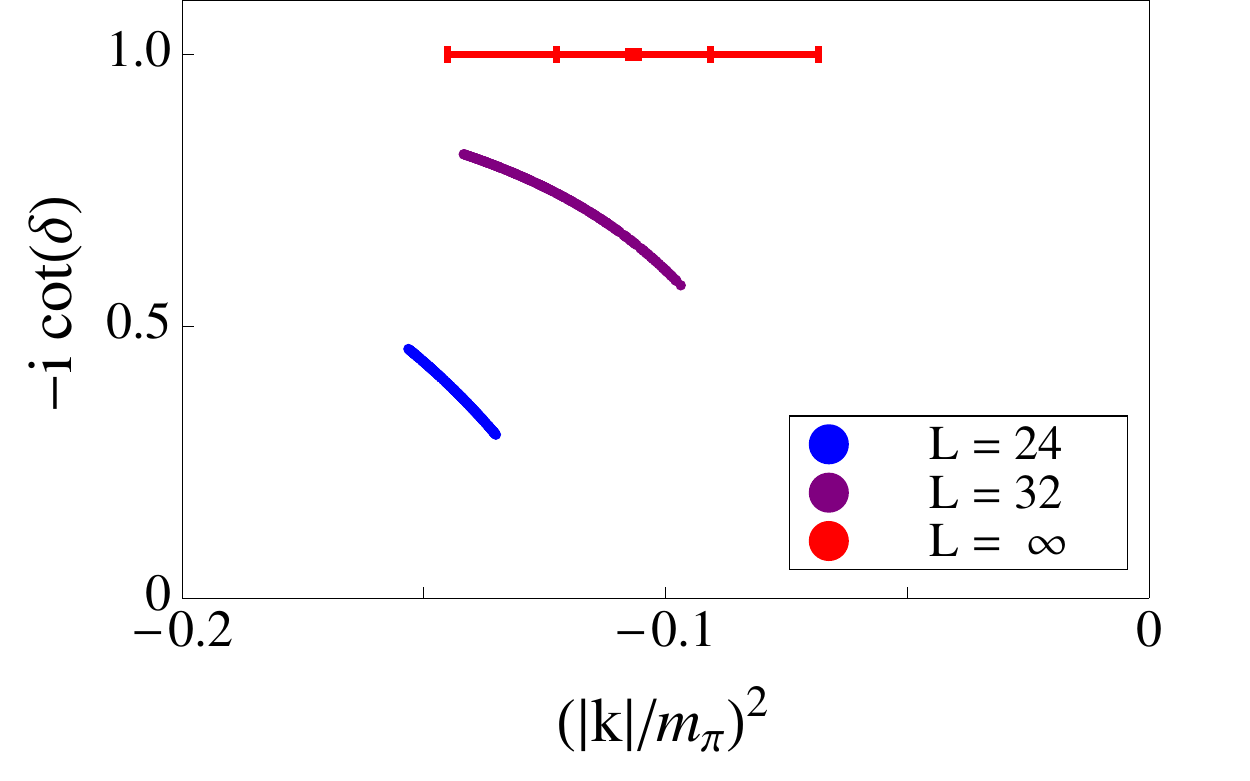}
\caption{The results of the LQCD calculations of $-i\cot\delta$ 
versus $|{\bf k}|^2/m_\pi^2$ 
in the H-dibaryon channel
obtained using eq.~(\protect\ref{eq:Luscher}),
along with the infinite-volume extrapolation using eq.~(\protect\ref{eq:pcotkappa}).
The statistical and systematic uncertainties have been combined in quadrature. 
}
  \label{fig:LLextrap}
\end{figure}

\begin{figure}[!ht]
  \centering
     \includegraphics[width=0.49\textwidth]{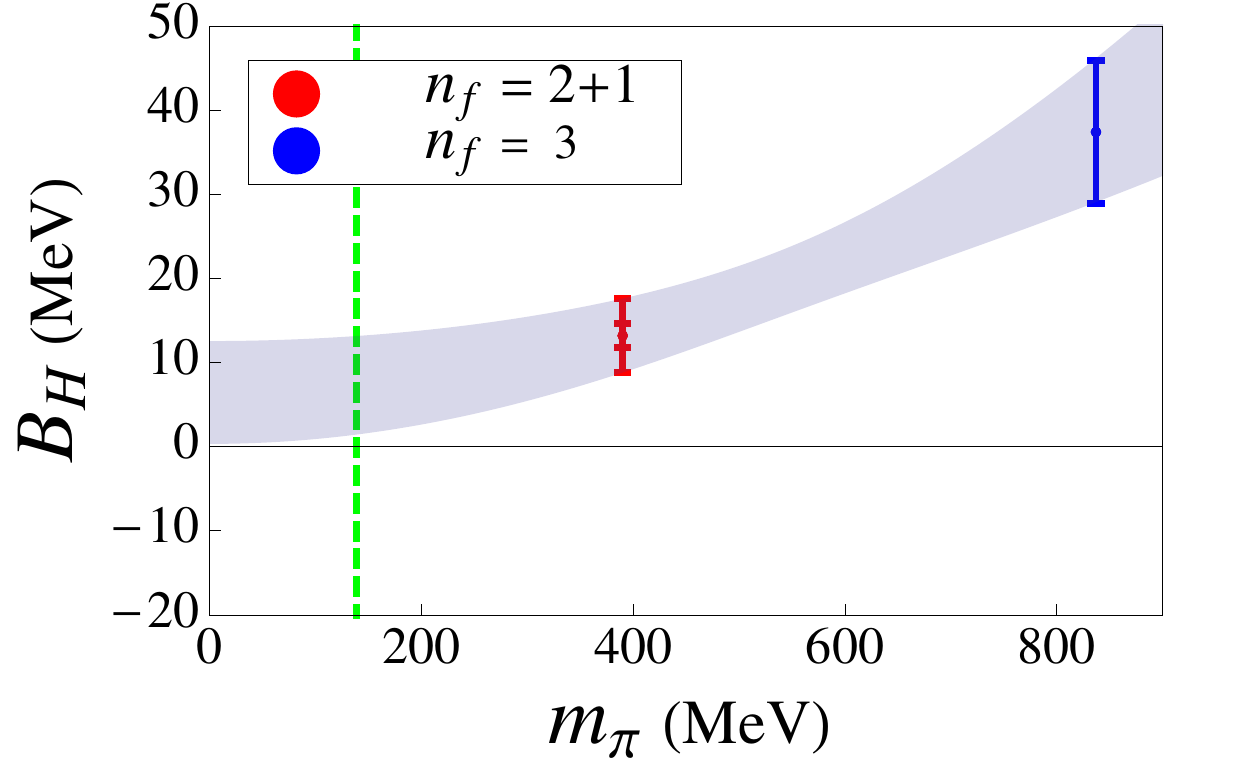}
     \includegraphics[width=0.49\textwidth]{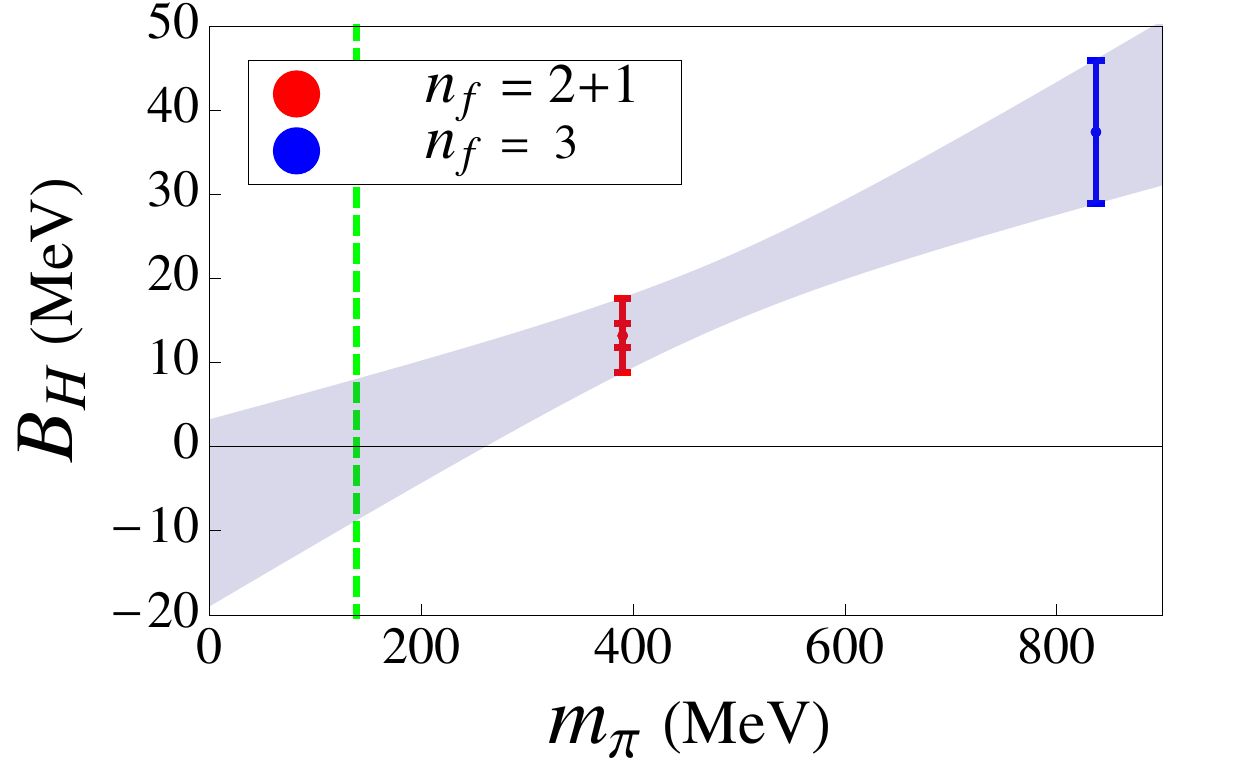}
\caption{Extrapolations of the LQCD results for the binding of the H-dibaryon.
The left panel corresponds to an extrapolation that is quadratic in $m_\pi$, of
the form $B_H(m_\pi) = B_0 + d_1 m_\pi^2$. The right panel is the same as a
left panel except with an extrapolation of the form $B_H(m_\pi) = \tilde B_0 +
\tilde d_1 m_\pi$.
In each panel, 
The blue point and uncertainty results from the $n_f=3$ LQCD
calculation of Ref.~\protect\cite{Inoue:2010es}, while the red point and
uncertainty is our present $n_f=2+1$ result.
The green dashed vertical line corresponds to the physical pion mass.
}
  \label{fig:Hextrap}
\end{figure}

In the absence of interactions, the $\Lambda\Lambda$-$\Xi N$-$\Sigma\Sigma$ 
coupled system (all three have the same quantum numbers)
is expected to exhibit three low-lying eigenstates
as the mass-splittings between the single-particle states 
are (from the $32^3\times 256$ ensemble)
\begin{eqnarray}
2(M_\Sigma - M_\Lambda) & = & 0.01317(13)(19)~{\rm t.l.u}
\ \ ,
\nonumber\\
M_\Xi + M_N - 2  M_\Lambda & = & 0.003397(61)(65)~{\rm t.l.u}
\ \ .
\label{eq:BBrestmasses}
\end{eqnarray}
However, if the interaction generates a bound state, it is unlikely
that a second or third state will also be bound, and therefore the
splitting between the ground state and the two additional states will
likely be larger than estimates based upon the single-baryon
masses.
The EMPs associated with the $\Lambda$ and the 
system with the quantum numbers of the $\Lambda\Lambda$
are shown in the left panels
of fig.~\ref{fig:Lamemp24} and fig.~\ref{fig:Lamemp32}. 
The binding energies extracted from the
LQCD calculations
are
\begin{eqnarray}
B_{H}^{(L=24)} & = & 
17.52\pm 0.88\pm 0.68~{\rm MeV}
\ \ ,\ \ 
B_{H}^{(L=32)}\ =\ 14.5\pm 1.3\pm 2.4~{\rm MeV}
\ \ \ ,
\label{eq:LLbindingLQCD}
\end{eqnarray}
which agree within uncertainties with the values given in our earlier
paper~\cite{Beane:2010hg}.
The volume extrapolation of the  results in  eq.~(\ref{eq:LLbindingLQCD})
is shown in fig.~\ref{fig:LLextrap}, and gives 
an extrapolated H-dibaryon binding energy of
\begin{eqnarray}
B_{H}^{(L=\infty)} & = & 
13.2\pm 1.8 \pm 4.0~{\rm MeV}
\ \ \ 
\label{eq:LLbindingLQCDextrap}
\end{eqnarray}
where the first uncertainty is statistical and the second is systematic.
In Ref.~\cite{Beane:2010hg},
$B_{H}^{(L=\infty)}$ was assigned a volume extrapolation
uncertainty of $\pm 1~{\rm MeV}$. In the present analysis, this systematic
uncertainty has been reduced to $\pm 0.3~{\rm MeV}$ by working to higher
orders in the volume expansion~\cite{Davoudi:2011md}.
The uncertainty in the energy-momentum relation is unchanged, and is estimated
to be $\pm 0.6~{\rm MeV}$. 
The updated result in eq.~(\ref{eq:LLbindingLQCDextrap})
at $m_\pi\sim 390~{\rm MeV}$ and the result of the
$n_f=3$ calculation at $m_\pi\sim 837~{\rm MeV}$~\cite{Inoue:2010es}
are shown in fig.~\ref{fig:Hextrap}.  Also shown in this figure are two
naive extrapolations, 
one that is linear in $m_\pi$ and one that is quadratic in $m_\pi$,
as discussed in Ref.~\cite{Beane:2011xf}.
The extrapolations indicate that the LQCD calculations are presently 
not at sufficiently small quark masses to determine
if the H-dibaryon is bound at the physical light-quark masses.

%%%%%%%%%%%%%%%%%%%%%%%%%%%%%%%%%%%%%%%%%%%%%%%%%%%%
\subsection{$\Xi^-\Xi^-$ }
\label{sec:XX}
\noindent
Experimental information on the hyperon-hyperon interactions in the $s < -2$ 
sector does not exist, 
presenting a significant handicap to studies of the composition of 
neutron star matter.
As an example of the importance of these interactions, 
Ref.~\cite{SchaffnerBielich:2000yj} shows that when
a strongly attractive $\Xi \Xi$ interaction is used in the 
Tolman-Oppenheimer-Volkoff equation,
new stable solutions appear, corresponding to compact hyperon stars with 
masses similar to neutron stars but with smaller radii.
From the theoretical point of view, 
the approximate flavor SU(3) symmetry of QCD
indicates that
a bound state 
in the $\Xi^-\Xi^-$ channel
is likely.
Phenomenological analyses of NN scattering and YN
scattering  provide a determination of the strength of the interaction for two
baryons in  the {\bf 27} irreducible representation of flavor
SU(3) that also contains the $\Xi^-\Xi^-$ system.
The OBE model developed by the Nijmegen group, NSC99~\cite{Stoks:1999bz}
\footnote{The recently developed extended soft-core models do not yet include
  the  $s < -2$ sectors.},
which include explicit breaking of flavor SU(3) symmetry by using the physical meson and baryon masses,
and chiral EFT~\cite{Savage:1995kv}, predicts  a 
bound state in the  $\Xi^- \Xi^-$ channel~\cite{Miller:2006jf,Haidenbauer:2009qn} at the physical pion mass~\footnote{
The $\Xi \Xi (\siii)$ and NN$(\siii)$ states belong to different irreducible
representations (${\bf 10}$ and $\overline{\bf 10}$,  respectively) and 
therefore SU(3) flavor symmetry alone 
is unable to predict whether 
an analog of the deuteron in the $s=-4$ sector exists. 
}.
However, only moderate attraction is obtained within a constituent quark model~\cite{Fujiwara:2006yh}.
A small $\Xi^- \Xi^-$ interaction was calculated in the $20^3 \times 128$
ensemble~\cite{Beane:2009py} used in this work
but may be subject to significant finite volume uncertainties. 
LQCD calculations performed in the flavor SU(3) limit~\cite{Inoue:2010hs},
in volumes of $16^3 \times 32$ with a lattice spacing of $b_s \sim 0.12~{\rm
  fm}$ and at pion masses of 1014 and 835 MeV 
found  attractive
interactions in the flavor singlet $t$-channel responsible for $\Xi^-\Xi^-$ interactions.

\begin{figure}[!ht]
  \centering
     \includegraphics[width=0.49\textwidth]{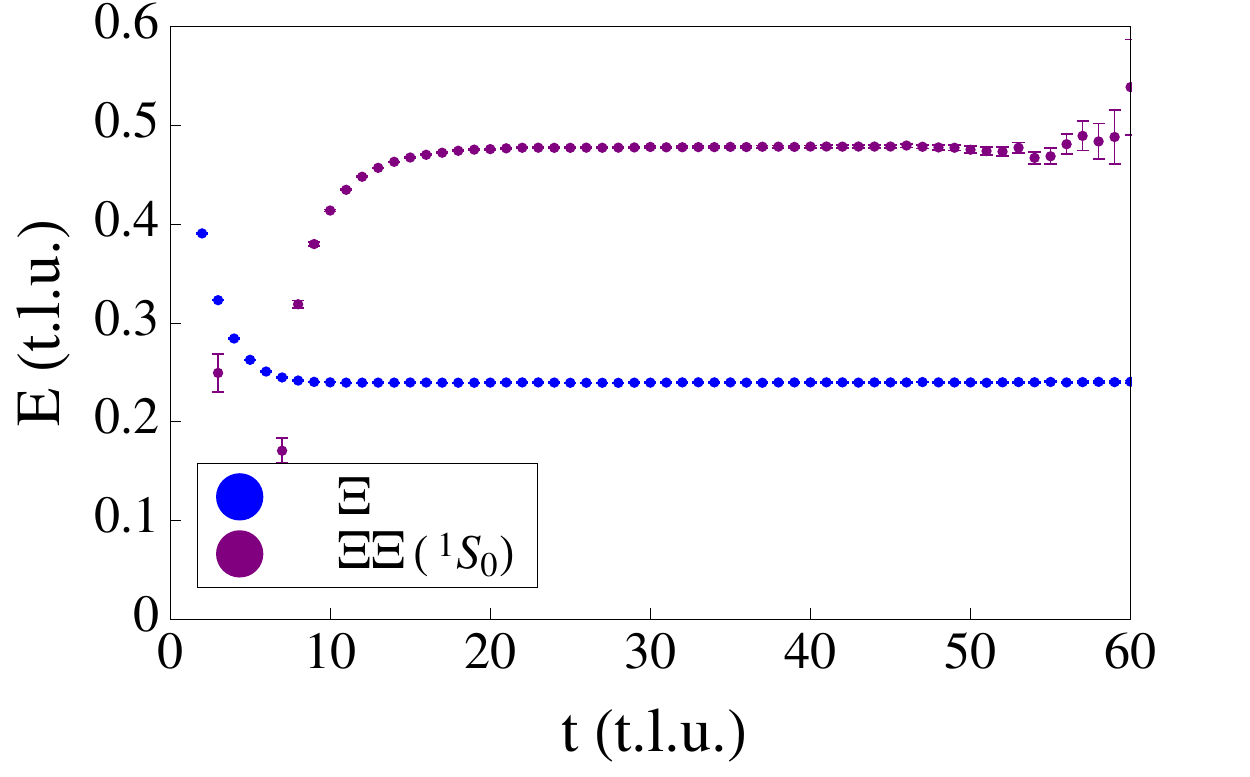}\ \  
     \includegraphics[width=0.49\textwidth]{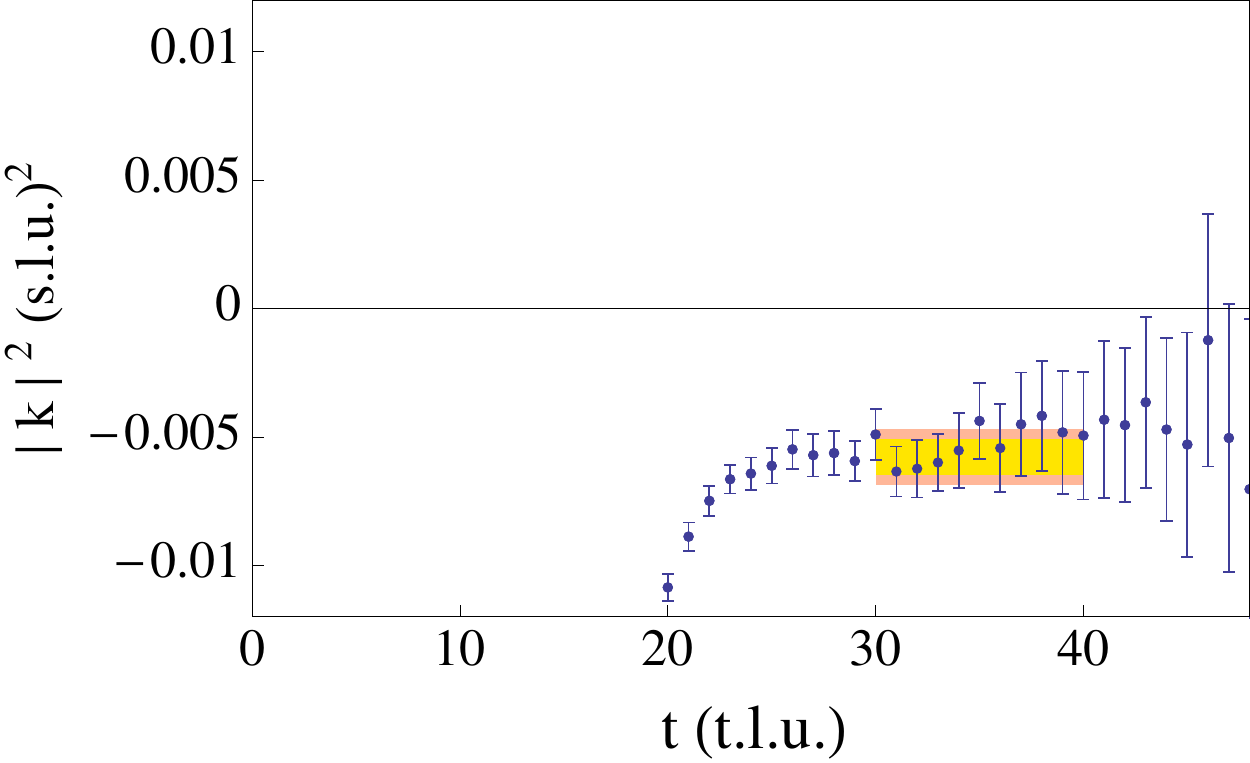}\ \  
\caption{The left panel shows an EMP of the $\Xi$ and of the $\Xi^-\Xi^-$
  system calculated with the $24^3\times
  128$ ensemble (in t.l.u.).  
The right panel shows the $|{\bf k}|^2$ (in $({\rm s.l.u.})^2$) of the
$\Xi^-\Xi^-$ system calculated with this ensemble, along with the fits.
}
  \label{fig:Xiemp24}
\end{figure}

\begin{figure}[!ht]
  \centering
     \includegraphics[width=0.49\textwidth]{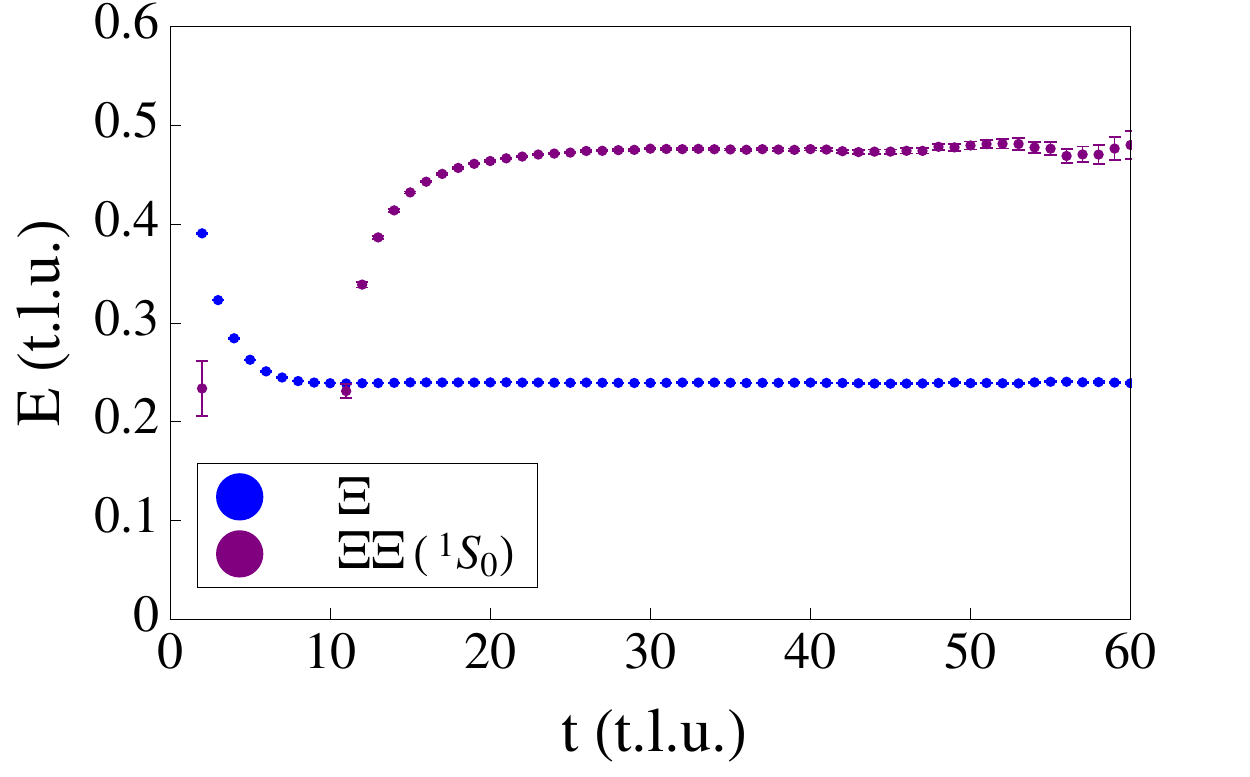}\ \  
     \includegraphics[width=0.49\textwidth]{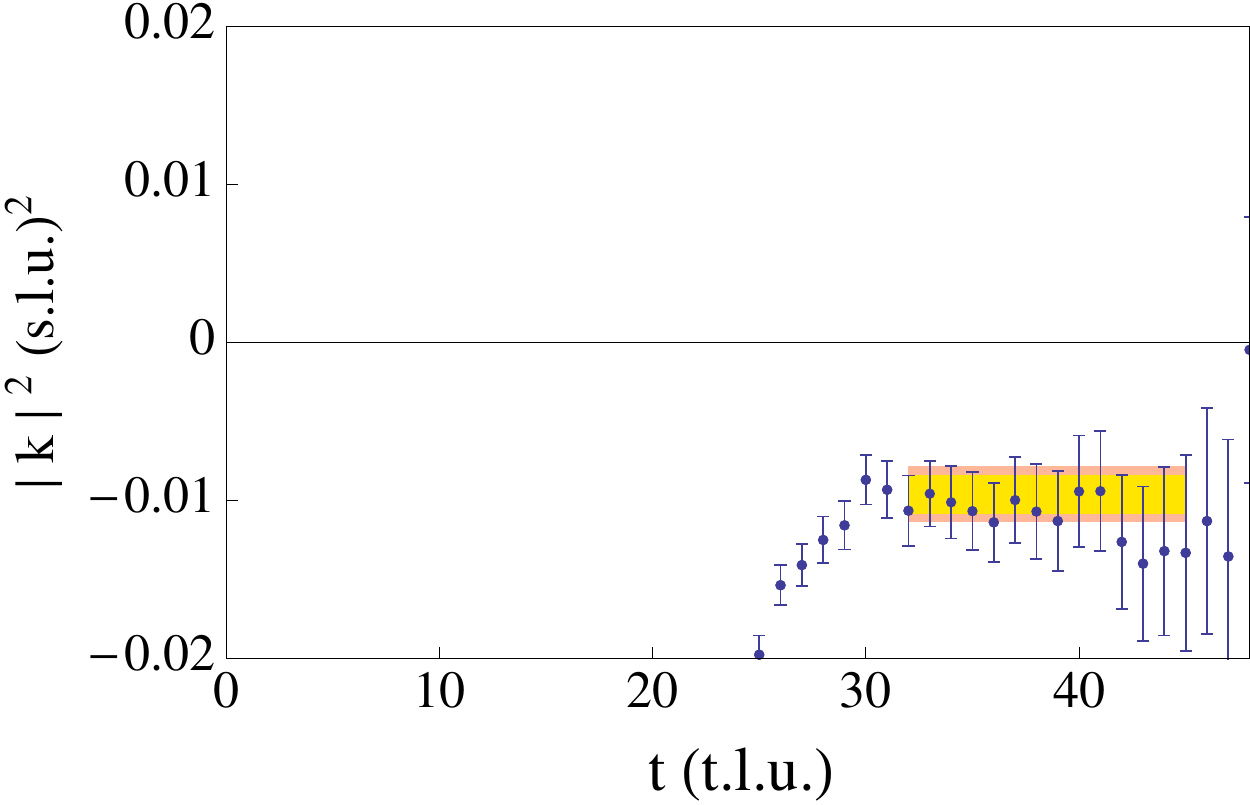}\ \  
\caption{The left panel shows an EMP of the $\Xi$ and of the $\Xi^-\Xi^-$
  system calculated with the 
$32^3\times 256$ ensemble (in t.l.u.).  
The right panel shows the $|{\bf k}|^2$ (in $({\rm s.l.u.})^2$) of the
$\Xi^-\Xi^-$ system calculated with this ensemble, along with the fits.
}
  \label{fig:Xiemp32}
\end{figure}

Our present LQCD calculations provide clear  evidence for a bound $\Xi^-\Xi^-$ state
at a pion mass of $m_\pi\sim 390~{\rm MeV}$.
The EMPs associated with the $\Xi$ and the
$\Xi^-\Xi^-$ system are shown in the left panels
of fig.~\ref{fig:Xiemp24} and fig.~\ref{fig:Xiemp32}. 

\begin{figure}[!ht]
  \centering
     \includegraphics[width=0.7\textwidth]{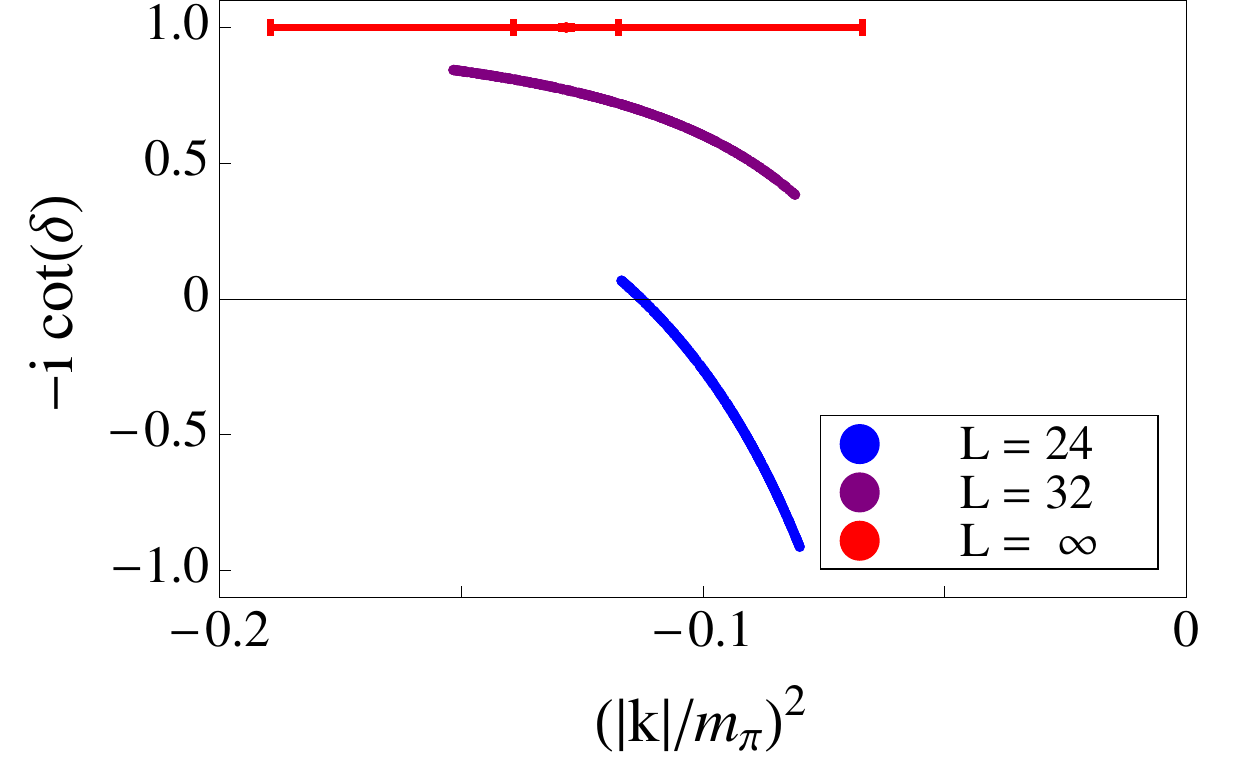}
\caption{The results of the Lattice QCD calculations of $-i\cot\delta$ 
versus $|{\bf k}|^2/m_\pi^2$ 
in the $\Xi^-\Xi^-$ system
obtained using eq.~(\protect\ref{eq:Luscher}),
along with the infinite-volume extrapolation using eq.~(\protect\ref{eq:pcotkappa}).
The inner uncertainty of the infinite-volume extrapolation is statistical,
while the outer corresponds to the statistical and systematic uncertainties
combined in quadrature. 
}
  \label{fig:XiXiextrap}
\end{figure}

The $\Xi^-\Xi^-$ binding energies extracted from the
LQCD calculations are 
\begin{eqnarray}
B_{\Xi^-\Xi^-}^{(L=24)} & = & 
11.0\pm 1.3\pm 1.6~{\rm MeV}
\ \ ,\ \ 
B_{\Xi^-\Xi^-}^{(L=32)}\ =\ 13.0\pm 0.5\pm 3.9~{\rm MeV}
\ \ \ .
\label{eq:XiXibindingLQCD}
\end{eqnarray}
The volume extrapolation of the  results in  eq.~(\ref{eq:XiXibindingLQCD})
is shown in fig.~\ref{fig:XiXiextrap}, and results in 
an extrapolated $\Xi^-\Xi^-$ binding energy of
\begin{eqnarray}
B_{\Xi^-\Xi^-}^{(L=\infty)} & = & 
14.0\pm 1.4 \pm 6.7~{\rm MeV}
\ \ \ 
\label{eq:XiXibindingLQCDextrap}
\end{eqnarray}
where the first uncertainty is statistical and the second is systematic.  
This indicates that, at the $\sim 2\sigma$ level,
the $\Xi^-\Xi^-$ channel supports a bound state.
The fact that the binding energy calculated in the $24^3\times 128$ ensemble
has $k\cot\delta\gsim 0$ indicates that this volume is 
significantly modifying
the  $\Xi^-\Xi^-$ bound state, and that calculations in larger
volumes, or with non-zero total momentum,
would refine the volume extrapolation.

\begin{figure}[!ht]
  \centering
     \includegraphics[width=0.7\textwidth]{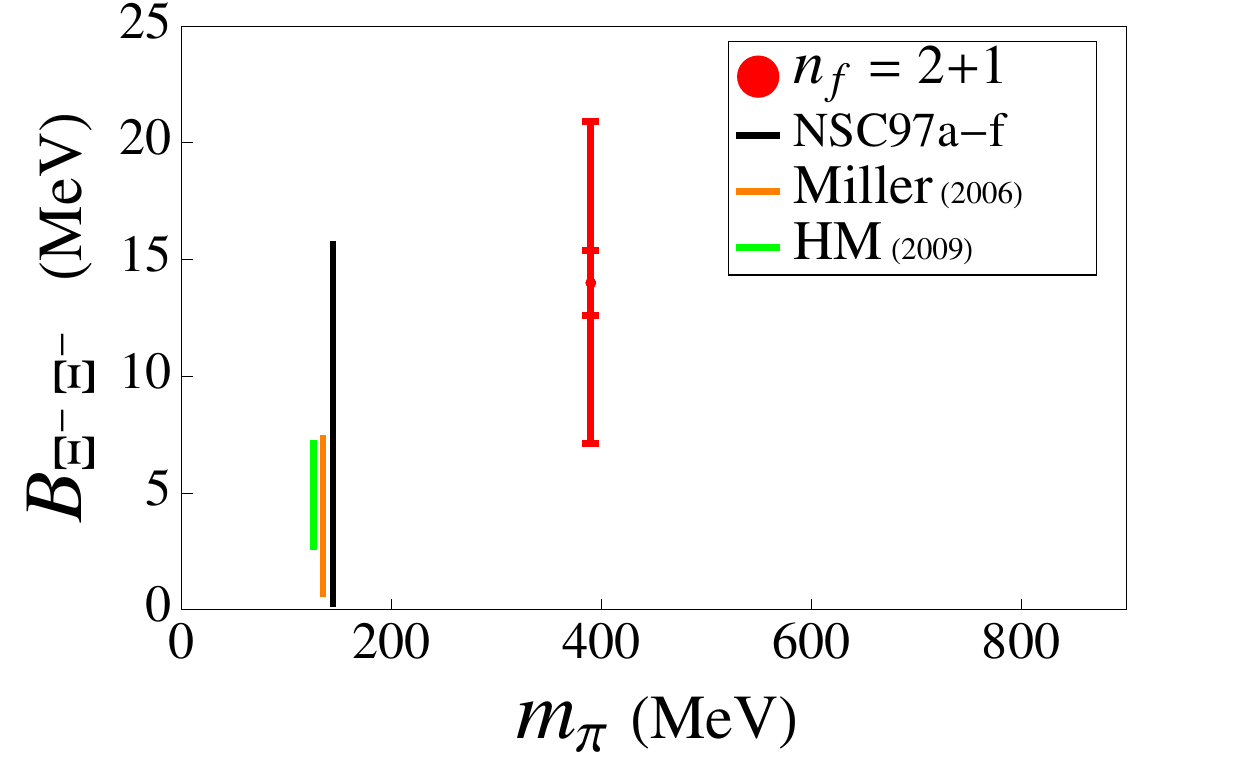}
\caption{The $\Xi^-\Xi^-$ binding energy as a function of the pion mass.
The black line denotes the predictions of the NSC97a-NSC97f  
models~\protect\cite{Stoks:1999bz} constrained from
nucleon-nucleon and hyperon-nucleon scattering data.
The orange line denotes the range of predictions by Miller~\cite{Miller:2006jf},
and the green line denotes the leading order EFT prediction by Haidenbauer and
Mei\ss ner (HM)~\cite{Haidenbauer:2009qn}.
The red point and uncertainty 
(the inner is statistical and the outer is statistical and systematic
combined in quadrature)
is our present $n_f=2+1$ result.
The OBE model and EFT predictions at the physical pion mass are displaced horizontally for
the purpose of display.
}
  \label{fig:XiXiALL}
\end{figure}
This result and the predictions of OBE models and leading order (LO) EFT are shown in
fig.~\ref{fig:XiXiALL}.
It is important to note that the uncertainty (and significance) of the LQCD
result is comparable to that of the OBE models and EFT results.
Further,
this result demonstrates that LQCD is rapidly approaching the situation
where it will provide more precise constraints on exotic systems
than can be achieved in the laboratory.
It will be interesting to see whether J-PARC~\cite{jparc} 
or FAIR~\cite{Steinheimer:2008hr}
can
provide constraints on the $s=-3$ and $s=-4$ systems, 
as well as on the possible H-dibaryon~\cite{jparcHdib}.
The binding energy 
in eq.~(\ref{eq:XiXibindingLQCDextrap})
provides strong motivation to return to OBE models and EFT frameworks
and determine the expected dependence on the light-quark masses.

%%%%%%%%%%%%%%%%%%%%%%%%%%%%%%%%%%%%%%%%%%%%%%%%%%%%
\subsection{$\Sigma^-\Sigma^-$ }
\label{sec:SS}
\noindent
As the $\Sigma^-\Sigma^-$ ($\si$) system is in the ${\bf 27}$ 
irreducible representation of flavor SU(3),
it is also expected to be bound, 
but by
somewhat less than 
the $\Xi^-\Xi^-$ system.
While the NSC97a-NSC97f models~\cite{Stoks:1999bz}
estimate the $\Sigma^- \Sigma^-$ binding, $B_{\Sigma^- \Sigma^-}$, to lie in the
range $1.5~{\rm MeV}\lsim B_{\Sigma^- \Sigma^-} \lsim 3.2$ MeV, 
large and negative scattering lengths are found in the $\Sigma^- \Sigma^-$
channel with LO EFT~\cite{Polinder:2007mp}
in the absence of Coulomb interactions and isospin breaking 
(these results exhibit non-negligible dependence on the
momentum cut-off). 
On the other hand, the constituent quark model of Ref.~\cite{Fujiwara:2006yh} finds 
strong similarities between the behavior of the $\Sigma^- \Sigma^-$ and 
$nn$ interactions, leading to similar values for the
phase shifts. 
Our LQCD calculations in this channel are inconclusive. While the ground state
in the $24^3\times 128$ ensemble is negatively shifted, 
the ground state in the $32^3\times 256$ ensemble is consistent
with zero, and thus is consistent with both a scattering state and a bound state.

%%%%%%%%%%%%%%%%%%%%%%%%%%%%%%%%%%%%%
\section{Conclusions}
\label{sec:Conclusions}
\noindent 
We have performed precise Lattice QCD calculations of 
baryon-baryon systems at a pion mass of  $m_\pi\sim 390~{\rm MeV}$ in
four ensembles of anisotropic Clover gauge-field configurations with a
spatial lattice spacing of $b_s\sim 0.123~{\rm fm}$, an anisotropy of
$\xi\sim 3.5$ and cubic spatial lattice volumes with extent $L\sim 2.0,
2.5, 3.0$ and $4.0~{\rm fm}$.
These calculations have provided evidence,  with varying levels of significance, for the
existence of two-baryon bound states from QCD, 
which are summarized in Table~\ref{tab:LQCDbinding}.
\begin{table}[!ht]
  \caption{
A summary of the two-body binding energies determined in this work.
  }
  \label{tab:LQCDbinding}
  \begin{ruledtabular}
    \begin{tabular}{c||ccccc}
        &   Deuteron  & Di-neutron & H-dibaryon & $\Xi^-\Xi^-$  \\
      \hline
 Binding Energy (MeV) & 11(05)(12) & 7.1(5.2)(7.3) & 13.2(1.8)(4.0) &
 14.0(1.4)(6.7) \\
    \end{tabular}
  \end{ruledtabular}
\end{table}
Our LQCD calculations were performed at one lattice spacing, $b_s\sim
0.123~{\rm fm}$, but discretization effects are expected to be small as they
scale as ${\cal O}\left(b_s^2\right)$ for the Clover action.
Consequently, we do not expect them to significantly alter our conclusions.
A second lattice spacing is required to quantify this systematic uncertainty.

By far the most significant result is that the H-dibaryon is bound at the
$3\sigma$ level at this pion mass, 
improving on results we have already 
presented in Ref.~\cite{Beane:2010hg}.
At the $\sim 2\sigma$ level of significance, we find that the $\Xi^-\Xi^-$
system is also bound, which is qualitatively consistent with an array of hadronic models and
EFT analyses of this system at the physical light-quark masses.
It is interesting  to note that the level of precision of 
the $\Xi^-\Xi^-$ binding from LQCD
is comparable to the level of precision associated with the phenomenological
predictions.  With increasing computational resources directed at these
two-baryon systems, the QCD prediction will become more precise and eventually
become input for phenomenological models and will be used to constrain the
coefficients appearing in the effective field theories.

A major goal of Lattice QCD is to postdict the
anomalously small
 binding energy of the
deuteron.
We have presented evidence for a bound deuteron from QCD, however the
$\sim 1\sigma$ level of significance is well
below ``discovery level'', and our result should be considered a first step toward a
definitive calculation.  
Nevertheless,
it is now unambiguously clear that 
a precise determination of the deuteron binding
energy can be performed with sufficient computational resources.
Our  result hints that the deuteron is
bound, as does the result of a previous quenched calculation, at heavy pion
masses, in contrast with phenomenological analyses and with 
EFT predictions.
We also find suggestions of a  bound di-neutron  which are far from definitive, 
but are consistent with the quenched result at a heavier pion
mass~\cite{Yamazaki:2011nd}. 
If this remains the
case when the calculation is refined, there are
light-quark masses between $m_\pi\sim 140~{\rm MeV}$ and $m_\pi\sim 390~{\rm MeV}$
for which the scattering length in this channel would be infinite and the system
would be scale-invariant at low energies.

Phenomenology based upon flavor SU(3) symmetry indicates that the $\Xi^-\Xi^-$ system
should be more bound than the $\Sigma^-\Sigma^-$ system, which in turn should be more
bound than the di-neutron (which is nearly bound) at the physical light-quark masses, as these three
systems are all members of the same ${\bf 27}$ irreducible representation of
SU(3).
Our results are consistent with this, but 
further work is required before definitive conclusions can be drawn.

The results of the Lattice QCD calculations presented in this paper, 
which  refine and broaden our previous work~\cite{Beane:2010hg},
provide clear evidence for bound-states of two baryons directly from
QCD. 
With the suggestion of a deuteron and a bound di-neutron at this heavier pion
mass, there is compelling  motivation to invest larger computational resources into 
pursuing Lattice QCD calculations at light-quark masses, and to perform such
calculations in multiple volumes and with multiple lattice spacings.
It is clear that enhanced computational resources will enable calculations of the properties
and interactions of nuclei from QCD with quantifiable and systematically
removable uncertainties.

\vskip0.2in

\noindent 
We would like to thank G. A. Miller for interesting discussions. 
We thank K. Roche for computing resources at ORNL NCCS, and 
B. Jo\'{o} for a significant contribution to our running.
We thank R. Edwards and B. Jo\'{o} for help with
QDP++ and Chroma~\cite{Edwards:2004sx}.  
We acknowledge computational
support from the USQCD SciDAC project, 
the National Energy Research Scientific Computing Center
(NERSC, Office of Science of the US
DOE, DE-AC02-05CH11231), the UW HYAK facility, Centro Nacional de
Supercomputaci\'on (Barcelona, Spain), LLNL, and the NSF through
Teragrid resources provided by TACC and NICS under grant number
TG-MCA06N025.  
SRB was supported in part by the NSF CAREER grant
PHY-0645570.  
The work of EC and
AP is supported by the contract FIS2008-01661 from MEC (Spain) and
FEDER.  
AP acknowledges support from the RTN Flavianet
MRTN-CT-2006-035482 (EU).  
H-WL and MJS were supported in part by the
DOE grant DE-FG03-97ER4014.  
WD and KO were supported in part by DOE
grants DE-AC05-06OR23177 (JSA) and DE-FG02-04ER41302.  
WD was also
supported by DOE OJI grant DE-SC0001784 and Jeffress Memorial Trust,
grant J-968.  
KO was also supported in part by NSF grant CCF-0728915 
and DOE OJI grant
DE-FG02-07ER41527.  
AT was supported by NSF grant PHY-0555234 and DOE
grant DE-FC02-06ER41443.  
The work of TL was performed under the
auspices of the U.S.~Department of Energy by LLNL under Contract
DE-AC52-07NA27344.  
The
work of AWL was supported in part by the Director, Office of Energy
Research, Office of High Energy and Nuclear Physics, Divisions of
Nuclear Physics, of the U.S. DOE under Contract No.  DE-AC02-05CH11231

%%%%%%%%%%%%%%%%%%%%%%%


\begin{thebibliography}{99}


% \cite{Jaffe:1976yi}
    \bibitem{Jaffe:1976yi} R.~L.~Jaffe,
  % ``Perhaps A Stable Dihyperon,''
  Phys.\ Rev.\ Lett.\ {\bf 38}, 195 (1977); {\bf 38},
  617 (1977)(E).
  %% CITATION = PRLTA,38,195;%%

%\cite{Stoks:1999bz}
\bibitem{Stoks:1999bz}
  V.~G.~J.~Stoks and T.~A.~Rijken,
  %``Soft core baryon baryon potentials for the complete baryon octet,''
  Phys.\ Rev.\  C {\bf 59}, 3009 (1999)
  [arXiv:nucl-th/9901028].
  %%CITATION = PHRVA,C59,3009;%%


%\cite{Miller:2006jf}
\bibitem{Miller:2006jf}
  G.~A.~Miller,
  %``Detecting Strangeness -4 Dibaryon States,''
  arXiv:nucl-th/0607006.
  %%CITATION = NUCL-TH/0607006;%%


%\cite{Haidenbauer:2009qn}
\bibitem{Haidenbauer:2009qn}
  J.~Haidenbauer, U.~-G.~Mei\ss ner,
  %``Predictions for the strangeness S = -3 and -4 baryon-baryon interactions in chiral effective field theory,''
  Phys.\ Lett.\  {\bf B684}, 275-280 (2010).
  [arXiv:0907.1395 [nucl-th]].



%\cite{Fukugita:1994na}
\bibitem{Fukugita:1994na}
  M.~Fukugita, Y.~Kuramashi, H.~Mino, M.~Okawa and A.~Ukawa,
  %``An Exploratory study of nucleon-nucleon scattering lengths in lattice
  %QCD,''
  Phys.\ Rev.\ Lett.\  {\bf 73}, 2176 (1994)
  [arXiv:hep-lat/9407012].
  %%CITATION = PRLTA,73,2176;%%

%\cite{Fukugita:1994ve}
\bibitem{Fukugita:1994ve}
  M.~Fukugita, Y.~Kuramashi, M.~Okawa, H.~Mino and A.~Ukawa,
  %``Hadron scattering lengths in lattice QCD,''
  Phys.\ Rev.\  D {\bf 52}, 3003 (1995)
  [arXiv:hep-lat/9501024].
  %%CITATION = PHRVA,D52,3003;%%

%\cite{Beane:2006mx}
\bibitem{Beane:2006mx}
  S.~R.~Beane, P.~F.~Bedaque, K.~Orginos and M.~J.~Savage,
  %``Nucleon-nucleon scattering from fully-dynamical lattice QCD,''
  Phys.\ Rev.\ Lett.\  {\bf 97}, 012001 (2006)
  [arXiv:hep-lat/0602010].
  %%CITATION = PRLTA,97,012001;%%

%\cite{Beane:2009py}
\bibitem{Beane:2009py}
  S.~R.~Beane {\it et al.}  [NPLQCD Collaboration],
  %``High Statistics Analysis using Anisotropic Clover Lattices: (III)
  %Baryon-Baryon Interactions,''
  Phys.\ Rev.\  D {\bf 81}, 054505 (2010)
  [arXiv:0912.4243 [hep-lat]].
  %%CITATION = PHRVA,D81,054505;%%


%\cite{Aoki:2008hh}
\bibitem{Aoki:2008hh}
  S.~Aoki, T.~Hatsuda and N.~Ishii,
  %``Nuclear Force from Monte Carlo Simulations of Lattice Quantum
  %Chromodynamics,''
  Comput.\ Sci.\ Dis.\  {\bf 1}, 015009 (2008)
  [arXiv:0805.2462 [hep-ph]].
  %%CITATION = COMSD,1,015009;%%

%\cite{Aoki:2009ji}
\bibitem{Aoki:2009ji}
  S.~Aoki, T.~Hatsuda and N.~Ishii,
  %``Theoretical Foundation of the Nuclear Force in QCD and its applications to
  %Central and Tensor Forces in Quenched Lattice QCD Simulations,''
  Prog.\ Theor.\ Phys.\  {\bf 123}, 89 (2010)
  [arXiv:0909.5585 [hep-lat]].
  %%CITATION = PTPKA,123,89;%%

%\cite{Ishii:2006ec}
\bibitem{Ishii:2006ec}
  N.~Ishii, S.~Aoki and T.~Hatsuda,
  %``The Nuclear Force from Lattice QCD,''
  Phys.\ Rev.\ Lett.\  {\bf 99}, 022001 (2007)
  [arXiv:nucl-th/0611096].
  %%CITATION = PRLTA,99,022001;%%

%\cite{Yamazaki:2011nd}
\bibitem{Yamazaki:2011nd}
  T.~Yamazaki, Y.~Kuramashi, A.~Ukawa,
  %``Two-Nucleon Bound States in Quenched Lattice QCD,''
  arXiv:1105.1418 [hep-lat].
  %%CITATION = ARXIV:1105.1418;%%

%\cite{Yamazaki:2009ua}
\bibitem{Yamazaki:2009ua}
  T.~Yamazaki, Y.~Kuramashi, A.~Ukawa,
  %``Helium Nuclei in Quenched Lattice QCD,''
  Phys.\ Rev.\  D {\bf 81}, 111504 (2010)
  [arXiv:0912.1383 [hep-lat]].
  %%CITATION = PHRVA,D81,111504;%%

%\cite{Beane:2009gs}
\bibitem{Beane:2009gs}
  S.~R.~Beane {\it et al.},
  %``High Statistics Analysis using Anisotropic Clover Lattices. II.
  %Three-Baryon Systems,''
  Phys.\ Rev.\  D {\bf 80}, 074501 (2009)
  [arXiv:0905.0466 [hep-lat]].
  %%CITATION = PHRVA,D80,074501;%%

%\cite{deForcrand:2009dh}
\bibitem{deForcrand:2009dh}
  P.~de Forcrand and M.~Fromm,
  %``Nuclear Physics from lattice QCD at strong coupling,''
  Phys.\ Rev.\ Lett.\  {\bf 104}, 112005 (2010)
  [arXiv:0907.1915 [hep-lat]].
  %%CITATION = PRLTA,104,112005;%%

%\cite{Beane:2010hg}
\bibitem{Beane:2010hg}
  S.~R.~Beane {\it et al.}  [NPLQCD Collaboration],
  %``Evidence for a Bound H-dibaryon from Lattice QCD,''
  Phys.\ Rev.\ Lett.\  {\bf 106}, 162001 (2011)
  [arXiv:1012.3812 [hep-lat]].
  %%CITATION = PRLTA,106,162001;%%

%\cite{Inoue:2010es}
\bibitem{Inoue:2010es}
  T.~Inoue {\it et al.}  [HAL QCD Collaboration],
  %``Bound H-dibaryon in Flavor SU(3) Limit of Lattice QCD,''
  Phys.\ Rev.\ Lett.\  {\bf 106}, 162002 (2011)
  [arXiv:1012.5928 [hep-lat]].
  %%CITATION = PRLTA,106,162002;%%

%\cite{Beane:2011xf}
\bibitem{Beane:2011xf}
  S.~R.~Beane {\it et al.},
  %``Present Constraints on the H-dibaryon at the Physical Point from Lattice
  %QCD,''
  arXiv:1103.2821 [hep-lat].
  %%CITATION = ARXIV:1103.2821;%%

%\cite{Shanahan:2011su}
\bibitem{Shanahan:2011su}
  P.~E.~Shanahan, A.~W.~Thomas and R.~D.~Young,
  %``Mass of the H-dibaryon,''
  arXiv:1106.2851 [nucl-th].
  %%CITATION = NONE,,;%%


%\cite{Beane:2009kya}
\bibitem{Beane:2009kya}
  S.~R.~Beane {\it et al.},
  %``High Statistics Analysis using Anisotropic Clover Lattices: (I) Single
  %Hadron Correlation Functions,''
  Phys.\ Rev.\  D {\bf 79}, 114502 (2009)
  [arXiv:0903.2990 [hep-lat]].
  %%CITATION = PHRVA,D79,114502;%%

%\cite{Beane:2011pc}
\bibitem{Beane:2011pc}
  S.~R.~Beane {\it et al.},
  %``High Statistics Analysis using Anisotropic Clover Lattices: (IV) Volume
  %Dependence of Light Hadron Masses,''
  Phys.\ Rev.\  D {\bf 84}, 014507 (2011)
  [arXiv:1104.4101 [hep-lat]].
  %%CITATION = PHRVA,D84,014507;%%





%%%%%%%%%%%   Luscher Method %%%%%%%%


  % \cite{Hamber:1983vu}
    \bibitem{Hamber:1983vu} H.~W.~Hamber, E.~Marinari, G.~Parisi and
  C.~Rebbi,
  % ``Considerations On Numerical Analysis Of QCD,''
  Nucl.\ Phys.\ B {\bf 225}, 475 (1983).
  %% CITATION = NUPHA,B225,475;%%

    \bibitem{Luscher:1986pf} M.~L\"uscher,
  % ``Volume Dependence of the Energy Spectrum in Massive Quantum
  % Field Theories.
  % 2. Scattering States,''
  Commun.\ Math.\ Phys.\ {\bf 105}, 153 (1986).
  %% CITATION = CMPHA,105,153;%%

    \bibitem{Luscher:1990ux} M.~L\"uscher,
  % ``Two particle states on a torus and their relation to the
  % scattering
  % matrix,''
  Nucl.\ Phys.\ B {\bf 354}, 531 (1991).
  %% CITATION = NUPHA,B354,531;%%


%\cite{Luscher:1985dn}
\bibitem{Luscher:1985dn}
  M.~L\"uscher,
  %``Volume Dependence of the Energy Spectrum in Massive Quantum Field Theories.
  %1. Stable Particle States,''
  Commun.\ Math.\ Phys.\  {\bf 104}, 177 (1986).
  %%CITATION = CMPHA,104,177;%%

%\cite{Beane:2003da}
\bibitem{Beane:2003da}
  S.~R.~Beane, P.~F.~Bedaque, A.~Parreno and M.~J.~Savage,
  %``Two nucleons on a lattice,''
  Phys.\ Lett.\  B {\bf 585}, 106 (2004)
  [arXiv:hep-lat/0312004].
  %%CITATION = PHLTA,B585,106;%%

%\cite{Bour:2011ef}
\bibitem{Bour:2011ef}
  S.~Bour, S.~Konig, D.~Lee, H.~W.~Hammer and U.~G.~Mei\ss ner,
  %``Topological phases for bound states moving in a finite volume,''
  arXiv:1107.1272 [nucl-th].
  %%CITATION = ARXIV:1107.1272;%%

%\cite{Davoudi:2011md}
\bibitem{Davoudi:2011md}
  Z.~Davoudi and M.~J.~Savage,
  %``Improving the Volume Dependence of Two-Body Binding Energies Calculated
  %with Lattice QCD,''
  arXiv:1108.5371 [hep-lat].
  %%CITATION = ARXIV:1108.5371;%%


%%%%%%%%%%   The details of the calc  %%%%%%%%%%%%%%%%

%\cite{Okamoto:2001jb}
 \bibitem{Okamoto:2001jb}
   M.~Okamoto {\it et al.}  [CP-PACS Collaboration],
   %``Charmonium Spectrum from Quenched Anisotropic Lattice QCD,''
   Phys.\ Rev.\  D {\bf 65}, 094508 (2002)
   [arXiv:hep-lat/0112020].
   %%CITATION = PHRVA,D65,094508;%%
 

%\cite{Chen:2000ej}
 \bibitem{Chen:2000ej}
   P.~Chen,
   %``Heavy quarks on anisotropic lattices: The charmonium spectrum,''
   Phys.\ Rev.\  D {\bf 64}, 034509 (2001)
   [arXiv:hep-lat/0006019].
   %%CITATION = PHRVA,D64,034509;%%

%\cite{Morningstar:2003gk}
\bibitem{Morningstar:2003gk}
  C.~Morningstar and M.~J.~Peardon,
  %``Analytic smearing of SU(3) link variables in lattice QCD,''
  Phys.\ Rev.\  D {\bf 69}, 054501 (2004)
  [arXiv:hep-lat/0311018].
  %%CITATION = PHRVA,D69,054501;%%

%\cite{Lin:2008pr}
 \bibitem{Lin:2008pr}
   H.~W.~Lin {\it et al.}  [Hadron Spectrum Collaboration],
   %``First results from 2+1 dynamical quark flavors on an anisotropic lattice:
   %light-hadron spectroscopy and setting the strange-quark mass,''
   Phys.\ Rev.\  D {\bf 79}, 034502 (2009)
   [arXiv:0810.3588 [hep-lat]].
   %%CITATION = PHRVA,D79,034502;%%
 

%\cite{Edwards:2008ja}
 \bibitem{Edwards:2008ja}
   R.~G.~Edwards, B.~Jo\'o and H.~W.~Lin,
   %``Tuning for Three-flavors of Anisotropic Clover Fermions with Stout-link
   %Smearing,''
   Phys.\ Rev.\  D {\bf 78}, 054501 (2008)
   [arXiv:0803.3960 [hep-lat]].
   %%CITATION = PHRVA,D78,054501;%%


%\cite{Dudek:2009qf}
 \bibitem{Dudek:2009qf}
   J.~J.~Dudek, R.~G.~Edwards, M.~J.~Peardon, D.~G.~Richards and C.~E.~Thomas,
   %``Highly excited and exotic meson spectrum from dynamical lattice QCD,''
   Phys.\ Rev.\ Lett.\  {\bf 103}, 262001 (2009)
   [arXiv:0909.0200 [hep-ph]].
   %%CITATION = PRLTA,103,262001;%%

 
%\cite{Bulava:2010yg}
\bibitem{Bulava:2010yg}
  J.~Bulava {\it et al.},
  %``Nucleon, $\Delta$ and $\Omega$ excited states in $N_f=2+1$ lattice QCD,''
  Phys.\ Rev.\  D {\bf 82} (2010) 014507
  [arXiv:1004.5072 [hep-lat]].
  %%CITATION = PHRVA,D82,014507;%%

%\cite{Dudek:2011tt}
\bibitem{Dudek:2011tt}
  J.~J.~Dudek, R.~G.~Edwards, B.~Joo, M.~J.~Peardon, D.~G.~Richards, C.~E.~Thomas,
  %``Isoscalar meson spectroscopy from lattice QCD,''
  Phys.\ Rev.\  {\bf D83}, 111502 (2011).
  [arXiv:1102.4299 [hep-lat]].


%\cite{Morningstar:2011ka}
\bibitem{Morningstar:2011ka}
  C.~Morningstar, J.~Bulava, J.~Foley, K.~J.~Juge, D.~Lenkner, M.~Peardon and C.~H.~Wong,
  %``Improved stochastic estimation of quark propagation with Laplacian
  %Heaviside smearing in lattice QCD,''
  Phys.\ Rev.\  D {\bf 83}, 114505 (2011)
  [arXiv:1104.3870 [hep-lat]].
  %%CITATION = PHRVA,D83,114505;%%

%\cite{Edwards:2011jj}
\bibitem{Edwards:2011jj}
  R.~G.~Edwards, J.~J.~Dudek, D.~G.~Richards, S.~J.~Wallace,
  %``Excited state baryon spectroscopy from lattice QCD,''
  [arXiv:1104.5152 [hep-ph]].


%\cite{Lin:2011da}
\bibitem{Lin:2011da}
  H.~-W.~Lin, S.~D.~Cohen,
  %``Roper Properties on the Lattice: An Update,''
  [arXiv:1108.2528 [hep-lat]].





%%%%%%%%%%%% Review  %%%%%%%%%%%%%%%%

%\cite{Beane:2010em}
\bibitem{Beane:2010em}
  S.~R.~Beane, W.~Detmold, K.~Orginos, M.~J.~Savage,
  %``Nuclear Physics from Lattice QCD,''
  Prog. \ Part.\ Nucl. \ Phys. {\bf 66}, 1, (2011)
    [arXiv:1004.2935 [hep-lat]].

%\cite{Bedaque:2006yi}
\bibitem{Bedaque:2006yi}
  P.~F.~Bedaque, I.~Sato and A.~Walker-Loud,
  %``Finite volume corrections to pi-pi scattering,''
  Phys.\ Rev.\  D {\bf 73}, 074501 (2006)
  [arXiv:hep-lat/0601033].
  %%CITATION = PHRVA,D73,074501;%%

%\cite{Sato:2007ms}
\bibitem{Sato:2007ms}
  I.~Sato and P.~F.~Bedaque,
  %``Fitting two nucleons inside a box: Exponentially suppressed corrections to
  %the Luscher's formula,''
  Phys.\ Rev.\  D {\bf 76}, 034502 (2007)
  [arXiv:hep-lat/0702021].
  %%CITATION = PHRVA,D76,034502;%%


\bibitem{Aubin:2010jc}
  C.~Aubin, K.~Orginos,
  %``A new approach for Delta form factors,'' 
  [arXiv:1010.0202 [hep-lat]].

\bibitem{ko}
K.~Orginos
%"Construction and Analysis of Two Baryon Correlation functions "
PoS(Lattice 2010)118 


%%%%%%%% deuteron   %%%%%%%%%%

%\cite{Beane:2002vs}
\bibitem{Beane:2002vs}
  S.~R.~Beane and M.~J.~Savage,
  %``Variation of fundamental couplings and nuclear forces,''
  Nucl.\ Phys.\  A {\bf 713}, 148 (2003)
  [arXiv:hep-ph/0206113].
  %%CITATION = NUPHA,A713,148;%%

%\cite{Epelbaum:2002gb}
\bibitem{Epelbaum:2002gb}
  E.~Epelbaum, U.~G.~Mei\ss ner and W.~Gloeckle,
  %``Nuclear forces in the chiral limit,''
  Nucl.\ Phys.\  A {\bf 714}, 535 (2003)
  [arXiv:nucl-th/0207089].
  %%CITATION = NUPHA,A714,535;%%

%\cite{Beane:2002xf}
\bibitem{Beane:2002xf}
  S.~R.~Beane and M.~J.~Savage,
  %``The quark mass dependence of two-nucleon systems,''
  Nucl.\ Phys.\  A {\bf 717}, 91 (2003)
  [arXiv:nucl-th/0208021].
  %%CITATION = NUPHA,A717,91;%%

%\cite{Chen:2010yt}
\bibitem{Chen:2010yt}
  J.~W.~Chen, T.~K.~Lee, C.~P.~Liu and Y.~S.~Liu,
  %``On the Quark Mass Dependence of Two Nucleon Observables,''
  arXiv:1012.0453 [nucl-th].
  %%CITATION = ARXIV:1012.0453;%%


%\cite{Flambaum:2007mj}
\bibitem{Flambaum:2007mj}
  V.~V.~Flambaum and R.~B.~Wiringa,
  %``Dependence of nuclear binding on hadronic mass variation,''
  Phys.\ Rev.\  C {\bf 76}, 054002 (2007)
  [arXiv:0709.0077 [nucl-th]].
  %%CITATION = PHRVA,C76,054002;%%

%\cite{Bedaque:2010hr}
\bibitem{Bedaque:2010hr}
  P.~F.~Bedaque, T.~Luu, L.~Platter,
  %``Quark mass variation constraints from Big Bang nucleosynthesis,''
  Phys.\ Rev.\  {\bf C83}, 045803 (2011).
  [arXiv:1012.3840 [nucl-th]].


%\cite{Cheoun:2011yn}
\bibitem{Cheoun:2011yn}
  M.~K.~Cheoun, T.~Kajino, M.~Kusakabe and G.~J.~Mathews,
  %``Time Dependent Quark Masses and Big Bang Nucleosynthesis Revisited,''
  arXiv:1104.5547 [astro-ph.CO].
  %%CITATION = ARXIV:1104.5547;%%




%%%%%%%%%%%%%%   di-neutron  %%%%%%%%%%%%%%

%\cite{Braaten:2003eu}
\bibitem{Braaten:2003eu}
  E.~Braaten, H.~W.~Hammer,
  %``An Infrared renormalization group limit cycle in QCD,''
  Phys.\ Rev.\ Lett.\  {\bf 91}, 102002 (2003).
  [nucl-th/0303038].


%%%%%%%%%%%%   H-dibaryon  %%%%%%%%%%%%%

%\cite{Bashinsky:1997qv}
\bibitem{Bashinsky:1997qv}
  S.~Bashinsky and R.~L.~Jaffe,
  %``Quark states near a threshold and the unstable H dibaryon,''
  Nucl.\ Phys.\  A {\bf 625}, 167 (1997)
  [arXiv:hep-ph/9705407].
  %%CITATION = NUPHA,A625,167;%%


%\cite{Yamamoto:2000wf}
\bibitem{Yamamoto:2000wf}
  K.~Yamamoto {\it et al.},
  %``Search for double-Lambda hypernuclei and the H dibaryon in the (K-,K+)
  %reaction on C-12,''
  Phys.\ Lett.\  B {\bf 478} (2000) 401.
  %%CITATION = PHLTA,B478,401;%%

%\cite{Sakai:1999qm}
\bibitem{Sakai:1999qm}
  T.~Sakai, K.~Shimizu and K.~Yazaki,
  %``H dibaryon,''
  Prog.\ Theor.\ Phys.\ Suppl.\  {\bf 137}, 121 (2000)
  [arXiv:nucl-th/9912063].
  %%CITATION = PTPSA,137,121;%%

%\cite{Mulders:1982da}
\bibitem{Mulders:1982da}
  P.~J.~Mulders and A.~W.~Thomas,
  %``Pionic Corrections And Multi - Quark Bags,''
  J.\ Phys.\ G {\bf 9}, 1159 (1983).
  %%CITATION = JPHGB,G9,1159;%%

%\cite{Trattner:2006jn}
\bibitem{Trattner:2006jn}
  A.~L.~Trattner, PhD Thesis, LBL, UMI-32-54109 (2006).
  %``Searching for the elusive H dibaryon,''
  %%CITATION = UMI-32-54109;%%

%\cite{Yoon:2007aq}
\bibitem{Yoon:2007aq}
  C.~J.~Yoon {\it et al.},
  %``Search for the H-dibaryon resonance in C-12 (K-, K+ Lambda Lambda X),''
  Phys.\ Rev.\  C {\bf 75}, 022201 (2007).
  %%CITATION = PHRVA,C75,022201;%%


%\cite{Mackenzie:1985vv}
\bibitem{Mackenzie:1985vv}
  P.~B.~Mackenzie and H.~B.~Thacker,
  %``Evidence Against A Stable Dibaryon From Lattice QCD,''
  Phys.\ Rev.\ Lett.\  {\bf 55}, 2539 (1985).
  %%CITATION = PRLTA,55,2539;%%


%\cite{Iwasaki:1987db}
\bibitem{Iwasaki:1987db}
  Y.~Iwasaki, T.~Yoshie and Y.~Tsuboi,
  %``THE H DIBARYON IN LATTICE QCD,''
  Phys.\ Rev.\ Lett.\  {\bf 60}, 1371 (1988).
  %%CITATION = PRLTA,60,1371;%%

%\cite{Pochinsky:1998zi}
\bibitem{Pochinsky:1998zi}
  A.~Pochinsky, J.~W.~Negele and B.~Scarlet,
  %``Lattice study of the H dibaryon,''
  Nucl.\ Phys.\ Proc.\ Suppl.\  {\bf 73}, 255 (1999)
  [arXiv:hep-lat/9809077].
  %%CITATION = NUPHZ,73,255;%%

%\cite{Wetzorke:1999rt}
\bibitem{Wetzorke:1999rt}
  I.~Wetzorke, F.~Karsch and E.~Laermann,
  %``Further evidence for an unstable H-dibaryon?,''
  Nucl.\ Phys.\ Proc.\ Suppl.\  {\bf 83}, 218 (2000)
  [arXiv:hep-lat/9909037].
  %%CITATION = NUPHZ,83,218;%%


%\cite{Wetzorke:2002mx}
\bibitem{Wetzorke:2002mx}
  I.~Wetzorke and F.~Karsch,
  %``The H dibaryon on the lattice,''
  Nucl.\ Phys.\ Proc.\ Suppl.\  {\bf 119}, 278 (2003)
  [arXiv:hep-lat/0208029].
  %%CITATION = NUPHZ,119,278;%%

%\cite{Luo:2007zzb}
\bibitem{Luo:2007zzb}
  Z.~H.~Luo, M.~Loan and X.~Q.~Luo,
  %``H-Dibaryon from Lattice QCD with Improved Anisotropic Actions,''
  Mod.\ Phys.\ Lett.\  A {\bf 22}, 591 (2007)
  [arXiv:0803.3171 [hep-lat]].
  %%CITATION = MPLAE,A22,591;%%



%\cite{Mondejar:2006yu}
\bibitem{Mondejar:2006yu}
  J.~Mondejar, J.~Soto,
  %``The nucleon-nucleon potential beyond the static approximation,''
  Eur.\ Phys.\ J.\  {\bf A32}, 77-85 (2007).
  [nucl-th/0612051].


%%%%%%% XiXi %%%%%%%%%%%%%

%\cite{SchaffnerBielich:2000yj}
\bibitem{SchaffnerBielich:2000yj}
  J.~Schaffner-Bielich, M.~Hanauske, H.~Stoecker and W.~Greiner,
  %``Hyperstars: Phase transition to (meta)stable hyperonic matter in neutron
  %stars,''
  arXiv:astro-ph/0005490.
  %%CITATION = ASTRO-PH/0005490;%%

%\cite{Savage:1995kv}
\bibitem{Savage:1995kv}
  M.~J.~Savage and M.~B.~Wise,
  %``Hyperon masses in nuclear matter,''
  Phys.\ Rev.\  D {\bf 53}, 349 (1996)
  [arXiv:hep-ph/9507288].
  %%CITATION = PHRVA,D53,349;%%

%\cite{Fujiwara:2006yh}
\bibitem{Fujiwara:2006yh}
  Y.~Fujiwara, Y.~Suzuki and C.~Nakamoto,
  %``Baryon-baryon interactions in the SU(6) quark model and their applications
  %to light nuclear systems,''
  Prog.\ Part.\ Nucl.\ Phys.\  {\bf 58}, 439 (2007)
  [arXiv:nucl-th/0607013].
  %%CITATION = PPNPD,58,439;%%

%\cite{Inoue:2010hs}
\bibitem{Inoue:2010hs}
  T.~Inoue {\it et al.}  [HAL QCD collaboration],
  %``Baryon-Baryon Interactions in the Flavor SU(3) Limit from Full QCD
  %Simulations on the Lattice,''
  Prog.\ Theor.\ Phys.\  {\bf 124}, 591 (2010)
  [arXiv:1007.3559 [hep-lat]].
  %%CITATION = PTPKA,124,591;%%


\bibitem{jparc}
{\tt http://j-parc.jp/NuclPart/index\_e.html}

%\cite{Steinheimer:2008hr}
\bibitem{Steinheimer:2008hr}
  J.~Steinheimer, M.~Mitrovski, T.~Schuster, H.~Petersen, M.~Bleicher and H.~Stoecker,
  %``Strangeness fluctuations and MEMO production at FAIR,''
  Phys.\ Lett.\  B {\bf 676}, 126 (2009)
  [arXiv:0811.4077 [hep-ph]].
  %%CITATION = PHLTA,B676,126;%%

\bibitem{jparcHdib}
{\tt http://j-parc.jp/NuclPart/pac\_1107/pdf/KEK\_J-PARC-PAC2011-03.pdf}


%%%%%%%%  sigma sigma  %%%%%%%%%%%
%\cite{Polinder:2007mp}
\bibitem{Polinder:2007mp}
  H.~Polinder, J.~Haidenbauer and U.~G.~Mei\ss ner,
  %``Strangeness S = -2 baryon-baryon interactions using chiral effective field
  %theory,''
  Phys.\ Lett.\  B {\bf 653}, 29 (2007)
  [arXiv:0705.3753 [nucl-th]].
  %%CITATION = PHLTA,B653,29;%%



\bibitem{Edwards:2004sx} R.~G.~Edwards and B.~Jo\'o,
  % ``The Chroma software system for lattice QCD,''
  Nucl.\ Phys.\ Proc.\ Suppl.\ {\bf 140} (2005) 832.
%  [arXiv:hep-lat/0409003].
  %% CITATION = NUPHZ,140,832;%%


\end{thebibliography}
\end{document}